\documentclass[a4paper, 12 pt]{article}
\usepackage[utf8]{inputenc}
\usepackage{jheppub}
\pdfoutput=1
\usepackage[english]{babel}
\usepackage{amsmath}
\usepackage{amssymb}
\usepackage{physics}
\usepackage{graphicx}
\usepackage{caption}
\usepackage{subcaption}
\usepackage{hyperref}
\usepackage{makecell}
\usepackage{multirow}
\usepackage{hhline}

\title{\vspace{-2cm}
	\begin{flushright}
		{\normalsize INR-TH-2024-013}
	\end{flushright}
	\vspace{0.5cm} Pauli form factor contributions to the inelastic proton bremsstrahlung \\ and dark photon production}

\author[a,b]{D. Gorbunov,}
\author[a,c]{E. Kriukova}

\affiliation[a]{Institute for Nuclear Research of the Russian Academy of Sciences, \\
	60th October Anniversary pr-ct 7a, Moscow 117312, Russia}
\affiliation[b]{Landau Phystech School of Physics and Research, Moscow Institute of Physics and Technology, \\
	Institutskiy per. 9, Dolgoprudny 141700, Russia}
\affiliation[c]{Faculty of Physics, Lomonosov Moscow State University,\\
	Leninskiye Gory 1-2, Moscow 119991, Russia}

\emailAdd{gorby@ms2.inr.ac.ru}
\emailAdd{kryukova.ea15@physics.msu.ru}

\abstract{We study the production of hypothetical vector particles, dark photons $\gamma^\prime$, with masses in the range 0.4--1.8\,GeV via inelastic proton bremsstrahlung. We further develop the approach of refs.~\cite{Gorbunov:2024,Mosphys,SessiyaOFNRAN}, where for the first time we considered the contributions to the cross section that are associated with the Pauli form factor in $pp\gamma^\prime$ vertex and obtained new splitting functions. We demonstrate numerically the importance of these corrections to full inelastic proton bremsstrahlung cross section and refine the sensitivity of the ongoing and future fixed-target experiments T2K, DUNE and SHiP to the parameters of dark photon model. A dedicated experiment on measurements of the proton electromagnetic form factors in the time-like region below the proton-antiproton threshold, like those suggested in PANDA at FAIR, would help to obtain the robust predictions for the dark photon production by a proton beam at fixed target.}
\arxivnumber{2409.11386}

\begin{document}
	\maketitle
	\flushbottom

\section{Introduction}
The discovery of neutrino oscillations, the observations of galaxy and galaxy clusters dynamics, measurements of  anisotropy and polarization of cosmic microwave background, the cosmic abundance of matter and the simultaneous absence of anti-matter clearly demonstrate the incompleteness of the Standard Model (SM). A universal and presently one of most popular ways to extend the SM is the portal framework~\cite{Agrawal:2021dbo}. This approach allows one to incorporate new physics particles, such as dark photons, dark scalars, axions and heavy neutral leptons, into the theory through various portals by adding to the SM Lagrangian the terms that are the products of operators composed of the SM and dark sector fields which respect both SM and dark sector gauge symmetries~\cite{Beacham:2019nyx}.

We consider the minimal dark photon model~\cite{Okun:1982xi,Galison:1983pa,Holdom:1985ag} (BC1 in Physics Beyond Colliders classification~\cite{Beacham:2019nyx}), in which the SM Lagrangian $\mathcal{L}_\text{SM}$ is extended by the vector field $\tilde{A}^{\prime}_\mu$, corresponding to the new gauge group $U(1)^\prime$,
\begin{equation}
	\mathcal{L}=\mathcal{L}_\text{SM}-\frac{1}{4}\tilde{F}^\prime_{\mu \nu}\tilde{F}^{\prime \mu \nu}-\frac{\epsilon}{2\cos\theta_W}\tilde{F}^\prime_{\mu\nu}B^{\mu \nu}+\frac{m_{\gamma'}^2}{2}\tilde{A}^\prime_\mu \tilde{A}^{\prime \mu},
\end{equation}
where $\tilde{F}^\prime_{\mu \nu}$ and $B_{\mu \nu}$ are the dark photon and SM hypercharge strength tensors correspondingly, $\epsilon$ is the kinetic mixing parameter, $\theta_W$ is the Weinberg mixing angle and the dark photon mass $m_{\gamma^\prime}$ can be originated e.g. from the Stueckelberg mechanism~\cite{Ruegg:2003ps}. By simultaneous rotation of the third component of the SM weak gauge field $W^3_\mu$, SM hypercharge field $B_\mu$ and the new vector field $\tilde{A}^\prime_\mu$, it is possible to diagonalize the kinetic and mass terms with an accuracy of $O(\epsilon)$~\cite{Miller:2021ycl}. This gives the new basis of vector fields: massive $Z$-boson, SM photon $A_\mu$ and dark photon $A^\prime_\mu$ with the same mass $m_{\gamma^\prime}$. Due to such rotation, the dark photon eventually couples to the electromagnetic SM current as 
\[
{\cal L}_\text{int}=-\epsilon e J_\text{em}^\mu A^{\prime}_\mu\,. 
\]

In similar scenarios the dark photon can also interact with dark sector fermions (including those forming the dark matter in ref.\,\cite{Pospelov:2007mp}), but in our study of dark photon production they do not play any role.  
Thus the model has only two essential parameters, the dark photon mass $m_{\gamma^\prime}$ and the kinetic mixing parameter $\epsilon$. The model parameter plane $(m_{\gamma^\prime},\epsilon)$ has been intensively explored\,\cite{Filippi:2020kii}. Moreover there are several ongoing and future experiments, such as T2K~\cite{deNiverville:2016rqh,Araki:2023xgb}, DUNE~\cite{DUNE:2020fgq,Breitbach:2021gvv} and SHiP~\cite{Gorbunov:2014wqa,SHiP:2015vad,SHiP:2020vbd}, that  plan to search for dark photons with masses around 1 GeV produced in $pp$-collisions by energetic protons on target.  

In this paper we study the production of dark photons with masses in range 0.4--1.8\,GeV in proton-proton collisions via the process of inelastic proton bremsstrahlung $pp\rightarrow \gamma^\prime X$ adopting the kinematics of a fixed-target experiment. In the considered mass region, the bremsstrahlung is the dominant production mode of $X$ in comparison with meson decays, e.g., $\pi^0,\eta\rightarrow\gamma\gamma^\prime$ and QCD Drell-Yan process $q\bar{q}\rightarrow\gamma^\prime$, which are more important at smaller and bigger dark photon masses $m_{\gamma^\prime}$ correspondingly.  

The paper is organized as follows. Section~\ref{sec:existing-res} contains a brief overview of the existing approaches to estimate the inelastic proton bremsstrahlung cross section. The description of proton electromagnetic and off-shell hadronic form factors used in calculations is presented in section~\ref{sec:form-factors}. Our main analytical results on inelastic bremsstrahlung cross section factorization taking into account both Dirac and Pauli proton electromagnetic form factors are outlined in section~\ref{sec:new-factorization}. We numerically compare our result for the full  cross section with previous estimates from the literature and present the analogous comparison for sensitivities of T2K, DUNE and SHiP experiments in sections~\ref{sec:numerical} and~\ref{sec:sensitivity}. Finally, section~\ref{sec:discuss} summarizes our findings.

\section{Inelastic proton bremsstrahlung: existing results} \label{sec:existing-res}
In this section we give an overview of the existing methods for estimating the cross section of inelastic proton bremsstrahlung $pp\rightarrow \gamma^\prime X$. Our study is largely based on the factorization procedure originally formulated by Altarelli and Parisi for the emission of massless quarks and gluons~\cite{Altarelli:1977zs}. Later it has been modified to describe the production of massive dark scalars via proton bremsstrahlung~\cite{Boiarska:2019jym} and has been further developed for the case of dark photon production by Foroughi-Abari and Ritz~\cite{Foroughi-Abari:2021zbm}. The latter method is briefly presented below. 

Figure\,\ref{fig:feyn-diag}
\begin{figure}[t]
	\begin{center}
		\includegraphics[width=0.4\textwidth]{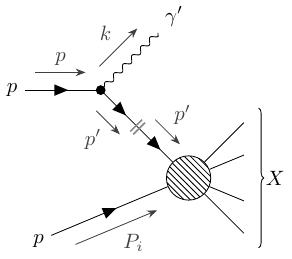}
		\caption{Dark photon production in the process of initial state radiation of inelastic proton bremsstrahlung.}
		\label{fig:feyn-diag}
	\end{center}
\end{figure}
shows the general Feynman diagram for the dark photon production via inelastic proton bremsstrahlung. Throughout this paper we denote the particles momenta as follows: $p$ is the momentum of the incident proton, $P_i$ is the momentum of the target proton, $k$ is the dark photon momentum, $p^\prime$ refers to the momentum of the virtual proton intermediate state. The $z$-axis of the Cartesian coordinate system in the lab frame is oriented along the 3-momentum of incident beam proton, so that the components of proton and dark photon momenta can be written as follows
\begin{align} \label{eq:momenta}
	p &= \{E_p, 0, 0, P\},\\
	k &= \{E_k, k_\perp \cos \varphi, k_\perp \sin \varphi, zP\},
\end{align}
where $P$ is the value of incident proton 3-momentum, $z$ is its fraction taken by the dark photon, $k_\perp$ is the transverse component of the dark photon 3-momentum. The energies of incident proton and dark photon are taken as
\begin{align}
	E_p &\equiv P + \frac{M^2}{2P},\\
	E_k &\equiv zP + \frac{m^2_{\gamma^\prime}+k^2_\perp}{2zP},
\end{align}
assuming that the second terms containing masses of proton $M$ and dark photon $m_{\gamma^\prime}$ are much smaller than $z$-components of their 3-momenta.

The cross section is factorized in the following way. At first, by presenting the proton propagator as the sum over spin states, one can extract from the matrix element of the full inelastic process $p(p)p(P_i)\rightarrow \gamma^\prime X$ the amplitude of the subprocess $p(p^\prime)p(P_i)\rightarrow X$ and the part responsible for dark photon emission (literally, one cuts the diagram in Fig.~\ref{fig:feyn-diag} along the double grey line). Then, using the explicit form of Dirac spinors and dark photon polarization vectors, one can obtain the splitting function $w_\text{FAR}(z, k^2_\perp)$, that is the probability density of dark photon emission, and connect the inelastic bremsstrahlung differential cross section with the full non-single diffractive $pp\rightarrow X$ cross section $\sigma_\text{NSD}(\bar{s})$ as
\begin{equation} \label{eq:FAR-result}
	\left[ \frac{\dd^2 \sigma (pp\rightarrow \gamma'X)}{\dd z \dd k^2_\perp}\right]_\text{FAR} = w_\text{FAR}(z, k_\perp^2)|F_1(m^2_{\gamma^\prime})|^2 F^2_\text{virt}(z, k^2_\perp) \sigma_\text{NSD}(\bar{s}),
\end{equation}
where $F_1(m^2_{\gamma^\prime})$ is the Dirac electromagnetic form factor of proton that comes from the $pp\gamma^\prime$ vertex due to the proton compositeness, $F_\text{virt}(z, k^2_\perp)$ is the off-shell hadronic form factor that controls the closeness of the intermediate proton to the mass shell. We comment further on the both form factors in section~\ref{sec:form-factors}. The splitting function by Foroughi-Abari and Ritz
\begin{equation} \label{eq:FAR-split-func}
	w_\text{FAR}(z, k^2_\perp)\equiv \frac{\epsilon^2 \alpha_\text{em}}{2\pi H} \left(z-\frac{z\left(1-z\right)}{H}\left(2M^2+m^2_{\gamma^\prime}\right)+\frac{H}{2zm^2_{\gamma^\prime}}\right)\\
\end{equation}
contains the fine-structure constant $\alpha_\text{em}$ and the common kinematic combination $H\equiv k^2_\perp+(1-z)m^2_{\gamma^\prime}+z^2M^2$. For the full non-single diffractive cross section we use the fit~\cite{Likhoded:2010pc}
\begin{equation}
	\sigma_\text{NSD}(s)=1.76+19.8\left(\frac{s}{\text{GeV}^2}\right)^{0.057}\text{\,mb},
\end{equation}
and take it at the square of the center-of-mass energy $s$ of intermediate and target protons equal to $\bar s\equiv 2 MP(1 - z) + 2 M^2 - H(z, k^2_\perp)/z$.

The approach described in~\cite{Foroughi-Abari:2021zbm} is clear, although we argue that it does \textit{not} give the \textit{complete} answer. The crucial thing is that the result~\eqref{eq:FAR-result} depends only on the Dirac electromagnetic proton form factor $F_1(m^2_{\gamma^\prime})$, although it should depend also on the analogous Pauli electromagnetic proton form factor $F_2(m^2_{\gamma^\prime})$, since they both contribute to the $pp\gamma^\prime$ vertex (see details in section~\ref{sec:form-factors}). In order to fully describe the inelastic proton bremsstrahlung process we extend the calculation from~\cite{Foroughi-Abari:2021zbm} and include the Pauli form factor in section~\ref{sec:new-factorization}. 

Another approximation for the inelastic bremsstrahlung cross section, which is the most frequently implemented for estimating the experimental sensitivity to dark photons, was developed by Blumlein and Brunner~\cite{Blumlein:2013cua}. Firstly, using the improved Weizsacker-Williams approximation~\cite{Kim:1973he}, the matrix element for inelastic process $pp\rightarrow\gamma^\prime p X$ is expressed via the amplitude of the $2\rightarrow 2$ process $p b \rightarrow\gamma^\prime p$, where $b$ is the auxiliary massless vector boson that mediates the $pp$-interaction. On the second step, the splitting probability for the subprocess $p\rightarrow \gamma^\prime p$ is derived. Finally, the differential cross section is factorized and presented as the product
\begin{equation} \label{eq:BB-result}
	\left[ \frac{\dd^2 \sigma (pp\rightarrow \gamma'pX)}{\dd z \dd k^2_\perp}\right]_{BB} = w_\text{BB}(z, k_\perp^2)|F_1(m^2_{\gamma^\prime})|^2\sigma_\text{inel}(s^\prime),
\end{equation}
where the splitting function in the Blumlein and Brunner approach is 
\begin{multline} \label{eq:BB-split-func}
	w_\text{BB}(z, k_\perp^2)\equiv\frac{\epsilon^2 \alpha_\text{em}}{2\pi H} \left[\frac{1+(1-z)^2}{z}-2z(1-z)\left(\frac{2M^2+m^2_{\gamma'}}{H}-z^2\frac{2M^4}{H^2}\right)+
	\right.\\\left. + 2z(1-z)(1+(1-z)^2)\frac{M^2 m_{\gamma'}^2}{H^2}+2z(1-z)^2\frac{m^4_{\gamma'}}{H^2}\right],
\end{multline}
and the inelastic proton scattering cross section $\sigma_\text{inel}$ is taken at $s^\prime$, which is the square of the center-of-mass energy of two protons that interact via the boson $b$. Unfortunately, it is unclear from \cite{Blumlein:2013cua}, how exactly the cross section~\eqref{eq:BB-result} was derived.

We should note that the expressions~\eqref{eq:FAR-split-func} and~\eqref{eq:BB-split-func} behave differently in several physically natural limits of the studied problem, such as the limits of small dark photon mass $m_{\gamma'} \ll M$, small momentum fraction $z \ll 1$, small (large) virtuality of the intermediate proton $M^2-(p-k)^2=H/z\ll M^2$ ($H/z\gg M^2$). It inevitably leads to noticeable discrepancies in the numerical results that have been earlier mentioned in~\cite{Foroughi-Abari:2021zbm} for inelastic bremsstrahlung and in~\cite{Gorbunov:2023jnx,Kriukova:2024wsi} for the similar case of elastic proton bremsstrahlung $pp\rightarrow pp\gamma^\prime$.

\section{Electromagnetic and off-shell hadronic form factors} \label{sec:form-factors}
The Dirac $F_1(t)$ and Pauli $F_2(t)$ electromagnetic proton form factors as the functions of squared transfer momentum $t$ are tightly connected with the matrix elements of the electromagnetic current
\begin{equation} \label{eq:emcurr}
	j^\mu_\text{em}\equiv\sum_i Q_i \bar{q_i} \gamma^\mu q_i,
\end{equation} 
that sums the inputs of light quarks $q_i \in \{u,d,s\}$ with electric charges $Q_i \in \{2/3, -1/3, -1/3\}$. Its matrix elements containing a proton with momentum $p_1$ in the initial state and a proton with momentum $p_2$ in the final state can be expressed through the electromagnetic proton form factors as follows
\begin{equation}
\bra{p(p_2)}j^\mu_\text{em}\ket{p(p_1)}\equiv \overline{u}(p_2) \left[F_1(t)\gamma^\mu + i\frac{F_2(t)}{2M}\sigma^{\mu\nu}(p_2-p_1)_\nu\right] u(p_1),	
\end{equation}
where $\sigma_{\mu \nu}\equiv i\left[\gamma_\mu, \gamma_\nu\right]/2$ and the squared transfer momentum $t\equiv (p_2-p_1)^2$. 

There are several different regions on the real $t$ axis. The space-like region is $t<0$, where the electromagnetic form factors are real and were measured in the process $ep\rightarrow ep$ of elastic electron scattering off proton e.g. by Jefferson Lab Hall A~\cite{JeffersonLabHallA:1999epl,JeffersonLabHallA:2001qqe,Puckett:2011xg}, MAMI and A1 collaboration~\cite{A1:2013fsc}.  The time-like region is $t>0$, where the form factors are complex-valued and the experimental investigations are possible only above the proton threshold $t>4M^2$. The time-like region has been recently explored by a number of experiments studying the creation of $p\bar p$ pairs in the $e^+e^-$ annihilation, among which are BESIII~\cite{BESIII:2021rqk}, BABAR~\cite{BaBar:2013ves}, CMD-3~\cite{CMD-3:2015fvi}, CLEO~\cite{CLEO:2005tiu}, etc. In between these two regions there is the so-called \textit{unphysical} region $0<t<4M^2$, that cannot be studied with the help of the reactions mentioned above, so the form factors in the unphysical region are usually obtained by means of interpolation or a fit which is valid in both space-like and time-like regions. Naturally one expects large and inevitable uncertainties in the values of electromagnetic form factors obtained in this way. Having nothing better, we adopt the estimates of these form factors from literature, since for our task it is this region in momentum transfer squared, where we need to describe the production of a dark photon of mass 0.4-1.8\,GeV. 

Let us also discuss the pros and cons of several available in literature estimates of the proton electromagnetic form factors in the unphysical region. An impressive fit based on the dispersion theory and both on space-like and time-like data from numerous experiments known up to day is presented in~\cite{Lin:2021xrc}. As an interesting feature, apart from resonances, this fit includes the continua, such as the two-pion contribution to the isovector part of the form factors derived in~\cite{Hoferichter:2016duk} and the $\rho \pi$ and $K\bar K$ continua inputs that contribute to isoscalar parts of the form factors~\cite{Lin:2021umz} (unfortunately, the last two ones are approximated as poles in the unphysical region). However, we could not make use of this fit in our work, since in order to reproduce the data it introduces a big number of fictitious zero-width resonances, which do not correspond to any known physical states, exactly in the unphysical region, that we want to investigate. Therefore it is not possible to predict the values of electromagnetic form factors in the vicinities of these poles even by introducing for them the non-zero decay widths (which in fact totally ruins the fit~\cite{Lin:2021xrc}).

Another well-known fit by Martemyanov, Faessler and Krivoruchenko~\cite{Faessler:2009tn}, based on the extended vector meson dominance model, is a common choice in most papers on proton bremsstrahlung. The fit utilizes photon mixing with $\rho$- and $\omega$-family mesons: each family contains one real $J^{PC}=1^{--}$ meson and two fictitious. Its advantages are the simple analytical form and the small number of free parameters. The Dirac ($j=1$) and Pauli ($j=2$) electromagnetic form factors of proton within this fit are parameterized in the similar way
\begin{equation} \label{eq:KrivFF}
	F_j(t)\equiv\sum_i \frac{f_{j,i}m^2_i}{m^2_i-t-im_i\Gamma_i},
\end{equation}
where the sum goes over all the six vector mesons with coupling constants to nucleons $f_{j,i}$, masses $m_i$ and decay widths $\Gamma_i$ listed in the table~\ref{tab:VMDconst}.
\begin{table}[t]
	\begin{center}
		\begin{tabular}{|c|c|c|c|c|c|c|}
			\hline
			& $\rho$ & $\omega$ & $\rho'$ & $\omega'$ & $\rho''$ & $\omega''$ \\
			\hline
			$f_{1,i}$ & 0.616 & 1.01 & 0.223 & -0.881 & -0.339 & 0.369 \\
			\hline
			$f_{2,i}$ & 4.17 & -0.143 & -4.42 & 0.151 & 2.11 & -0.723 \\
			\hline
			$m_i$,\,GeV & \multicolumn{2}{c|}{0.770} & \multicolumn{2}{c|}{1.25} & \multicolumn{2}{c|}{1.45} \\
			\hline
			$\Gamma_i$,\,GeV & 0.150 & 0.0085 & \multicolumn{2}{c|}{0.300} & \multicolumn{2}{c|}{0.500} \\
			\hline
		\end{tabular}    
	\end{center}
	\caption{The parameters of $\rho$- and $\omega$-family mesons used in the fit~\cite{Faessler:2009tn} for the proton Dirac and Pauli electromagnetic form factors~\eqref{eq:KrivFF}.}
	\label{tab:VMDconst}
\end{table}
Unfortunately, this parameterization did not agree with experimental data in the time-like region that were available by the time of publication~\cite{Faessler:2009tn}. Moreover, in the recent 10 years  even more new experimental data have been obtained in the time-like region that, of course, were not included in the study~\cite{Faessler:2009tn}. There are also some theoretical drawbacks: the continua input is not taken into account at all and the parameters of the resonances $\rho^\prime$, $\rho^{\prime \prime}$, $\omega^\prime$, $\omega^{\prime \prime}$ are completely artificial, which can be safe when one considers the form factor in the space-like or in the physical time-like region, but seriously affects the form factors values near $t=m^2_i$ in the region of our interest. We also doubt the model unitarity, since the fit considers the sum of overlapping Breit-Wigner resonances instead of working in the framework of $\mathcal{K}$-matrix formalism~\cite{Aitchison:1972ay}.

As a promising alternative we consider the fits, obtained in works of Dubnicka, Dubnickova and coauthors~\cite{Dubnicka:2002yp,Adamuscin:2016rer,Dubnickova:2020heq}. The fits are based on the unitary and analytic model for nucleon electromagnetic form factors~\cite{Dubnicka:2002yp}. The most recent fit built on modern experimental data~\cite{Dubnickova:2020heq} contains the impacts of nine experimentally-confirmed neutral vector mesons $\rho(770)$, $\omega(782)$, $\phi(1020)$, $\rho^\prime(1450)$, $\omega^\prime(1420)$, $\phi^\prime(1680)$, $\rho^{\prime \prime}(1700)$, $\omega^{\prime \prime}(1650)$, $\phi^{\prime \prime}(2170)$ with quantum numbers $1^{--}$ and experimentally measured masses and decay widths. The form factors are presented in quite complicated analytical form that we, actually, failed to reproduce using the expressions from~\cite{Dubnickova:2020heq}. Thus in this paper we worked with the tabulated proton electromagnetic form factors that were helpfully provided by the authors. One should also note that there exist an observed disagreement between the total cross section of electron-positron annihilation to $p\bar p$ measured by BESIII collaboration and the theoretical prediction~\cite{Dubnickova:2020heq} made using these form factors. 

The electromagnetic proton form factors discussed above were obtained assuming that both protons are on the mass shell. However, in the process of proton bremsstrahlung~(see figure~\ref{fig:feyn-diag}) only the incident proton is on-shell. The general theoretical consideration of such half off-shell matrix elements of the electromagnetic current $j^\mu_\text{em}$ and their connection to the electromagnetic Dirac $F_1(t)$ and Pauli $F_2(t)$ form factors (unfortunately, without any concrete numerical estimates of changes in the form factors values) was presented in~\cite{Haberzettl:2011zr}. Moreover, there are investigations of extended pion electromagnetic form factors from the subprocess $\pi\rightarrow \gamma^* \pi^*$, that depend on two parameters: the photon momentum squared $q^2$ and the off-shell pion momentum squared $t$~\cite{Choi:2019nvk,Leao:2024agy}. These form factors were inferred both from experimental data on the exclusive cross section $ep\rightarrow e \pi^+ n$ and using the exactly solvable manifestly covariant model. As function of $t$ they rapidly decrease for pions moving away from the mass shell $t=m^2_{\pi}$. Similarly, we assume that proton electromagnetic form factors also rapidly decrease for protons departing from the mass shell and therefore we introduce the conservative phenomenological off-shell hadronic form factor~\cite{Feuster:1998cj}
\begin{equation} \label{eq:off-shell-FF}
	F_\text{virt}(z, k^2_\perp)\equiv \frac{\Lambda^4}{\Lambda^4+H^2(z, k^2_\perp)/z^2}
\end{equation}
with the scale $\Lambda=1.5$\,GeV that suppresses the matrix elements for intermediate protons with large virtuality $(p-k)^2-M^2=-H/z$. 

To sum up, in this work for numerical calculations of the cross sections in section~\ref{sec:numerical} and for revising the experimental sensitivities in section~\ref{sec:sensitivity} we exploit the fits for proton electromagnetic form factors from Refs.\,\cite{Faessler:2009tn,Dubnickova:2020heq} and the off-shell hadronic form factor~\eqref{eq:off-shell-FF}. We found that the numerical results depend on both electromagnetic and hadronic form factors. Replacing the latter with unity increases the corresponding number of signal events by a factor (different for different experiments and dark photon masses) of several, which means that more sophisticated account for proton virtually is more than welcome. The choice of electromagnetic form factors is described and its impact on the sensitivity is explored in details in sections \ref{sec:numerical} and \ref{sec:sensitivity}. 

\section{Cross section factorization} \label{sec:new-factorization}

In this section we present the detailed calculation of the inelastic proton bremsstrahlung $pp\rightarrow \gamma^\prime X$ differential cross section. The results have been presented for the first time in refs.~\cite{Gorbunov:2024,Mosphys,SessiyaOFNRAN} with only a brief explanation. Relying on the results obtained in~\cite{Alterelli:1964bu,Baier:1966jf,BERESTETSKII1982354} for photons in the electron bremsstrahlung, we neglect the dark photon production by the target proton. We follow~\cite{Foroughi-Abari:2021zbm} in acknowledging the quasi-real approximation, where both protons, incident and target, dissociate after scattering and the intermediate proton is close to the mass shell, and estimate the inelastic bremsstrahlung only as the initial state radiation depicted in fig.~\ref{fig:feyn-diag}. Here we extend the approach of ref.\,\cite{Foroughi-Abari:2021zbm} and consider the full impact of $pp\gamma^\prime$ vertex to the dark photon production, containing both terms with Dirac $F_1(t)$ and Pauli $F_2(t)$ electromagnetic form factors.   

In contrast to ref.\,\cite{Boiarska:2019jym}, where the old-fashioned perturbation theory is exploited, we work in the framework of usual modern quantum field theory and hence we have the energy-momentum conservation in every vertex. Thus in addition to~\eqref{eq:momenta} we have 
\begin{equation}
	p^\prime = \{E_{p^\prime}, -k_\perp \cos \varphi, -k_\perp \sin \varphi, (1-z)P\},
\end{equation}
where the intermediate proton energy reads $E_{p^\prime}\equiv E_p-E_k$. We would like to cast the differential cross section to a product 
\begin{equation} \label{eq:our-wanted-result}
	\frac{\dd^2 \sigma(pp\rightarrow \gamma^\prime X)}{\dd z \dd k^2_\perp}\simeq w_\text{mas}(z, k^2_\perp) F^2_\text{virt}(z, k^2_\perp) \sigma_\text{NSD}(\bar{s}),
\end{equation}
where the master splitting function $w_\text{mas}(z, k^2_\perp)$ depends on the quadratic combinations of Dirac and Pauli electromagnetic form factors. While deriving~\eqref{eq:our-wanted-result} we omit the phenomenological hadronic off-shell form factor $F_\text{virt}(z, k^2_\perp)$ for simplicity, but restore it at the very end.

First, we extract the matrix element $\mathcal{M}^{r^\prime}\equiv A(p-k, P_i)u^{r^\prime}(p-k)$ of the inelastic $pp$-scattering subprocess $p(p-k)p(P_i)\rightarrow X$ from the full matrix element of $p(p)p(P_i)\rightarrow \gamma^\prime(k)X$, which we denote as $\mathcal{M}^{r\lambda}$. Here $r\text{, }r^\prime \in \{\uparrow, \downarrow\}$ are the spinor indices numerating the spin states of the incident and intermediate protons, $\lambda\in \{+,-,L\}$ is one of dark photon polarizations and we omit the polarization of the target proton, since it does not affect the calculations. We start from expression for the full matrix element   
\begin{multline} \label{eq:Mrl}
	\mathcal{M}^{r\lambda}=A(p-k, P_i)\frac{i(\hat{p}-\hat{k}+M)}{(p-k)^2-M^2}(-i\epsilon e)\times\\\times\left(\gamma_\mu F_1\left(m^2_{\gamma^\prime}\right)+\frac{i}{2M}\sigma_{\mu \nu}(-k^\nu)F_2\left(m^2_{\gamma^\prime}\right)\right)(\epsilon^\lambda)^{*,\mu}(k)u^r(p),
\end{multline}
where $e$ is the proton electric charge and electromagnetic form factors are taken at $k^2=m^2_{\gamma^\prime}$. Next, we decompose the numerator of proton propagator as a sum over spin states,
\begin{equation}
	\hat{p}-\hat{k}+M = \sum_{r^\prime} u^{r^\prime}(p-k)\overline{u}^{r^\prime}(p-k),
\end{equation}
and simplify the intermediate proton virtuality in the denominator as 
\begin{equation}
	(p-k)^2-M^2=-H/z,
\end{equation}
which gives
\begin{equation} \label{eq:Mrpr}
	\mathcal{M}^{r\lambda}=-\sum_{r^\prime} \mathcal{M}^{r^\prime} \frac{\epsilon ez}{H} \left(V_1^{r^\prime r \lambda} F_1\left(m^2_{\gamma^\prime}\right)+V_2^{r^\prime r \lambda} F_2\left(m^2_{\gamma^\prime}\right)\right),
\end{equation}
where we introduced the vertex functions
\begin{align} 
	V_1^{r^\prime r \lambda} &\equiv \overline{u}^{r^\prime}(p-k) \widehat{(\epsilon^\lambda)^*} u^r(p), \label{eq:vertex-func-1}\\
	V_2^{r^\prime r \lambda} &\equiv \frac{1}{4M} \overline{u}^{r^\prime}(p-k)\left[\widehat{(\epsilon^\lambda)^*}, \hat{k}\right] u^r(p). \label{eq:vertex-func-2}
\end{align}

Second, we find the explicit form of the vertices in terms of the kinematic variables. The vertex function \eqref{eq:vertex-func-1} is the only one earlier considered in ref.\,\cite{Foroughi-Abari:2021zbm}, while \eqref{eq:vertex-func-2} corresponds to the novel term in~\eqref{eq:Mrpr} multiplied by the Pauli electromagnetic form factor. To obtain their explicit form, we substitute the spinors in the Dirac representation
\begin{align}
	u_\uparrow(p) &= \left(\sqrt{E_p+M}, 0, \frac{P}{\sqrt{E_p+M}}, 0\right)^T, \\
	u_\downarrow(p) &= \left(0, \sqrt{E_p+M}, 0, \frac{-P}{\sqrt{E_p+M}}\right)^T, \\
	u_\uparrow(p^\prime) &= \sqrt{E_{p^\prime}+M} \left(1, \frac{-k_\perp e^{i\varphi}}{2(1-z)P}, \frac{(1-z)P}{E_{p^\prime}+M}, \frac{-k_\perp e^{i\varphi}}{2(E_{p^\prime}+M)}\right)^T, \\
	u_\downarrow(p^\prime) &= \sqrt{E_{p^\prime}+M} \left(\frac{-k_\perp}{2(1-z)P}, -e^{i\varphi}, \frac{k_\perp}{2(E_{p^\prime}+M)}, \frac{(1-z)Pe^{i\varphi}}{E_{p^\prime}+M}\right)^T
\end{align}
and the dark photon polarization vectors
\begin{align}
		\epsilon^\mu_+&=\frac{1}{\sqrt{2}}\left\{0, 1, i, \frac{-k_\perp}{zP}e^{i\varphi}\right\},\\
		\epsilon^\mu_-&=\frac{1}{\sqrt{2}}\left\{0, 1, -i, \frac{-k_\perp}{zP}e^{-i\varphi}\right\},\\
		\epsilon^\mu_L&=\frac{1}{m_{\gamma^\prime}} \left\{zP+\frac{k^2_\perp}{2zP}, k_\perp \cos \varphi, k_\perp \sin \varphi, zP+\frac{m^2_{\gamma^\prime}}{2zP}\right\}
\end{align}
into expressions~\eqref{eq:vertex-func-1}, \eqref{eq:vertex-func-2}. This gives the following vertex functions to the leading (not greater than the second) order in $k_\perp$, $M$, $m_{\gamma^\prime}$ for the circular dark photon polarizations, $\lambda=\pm$, 
\begin{align}
	V^{r^\prime r\lambda}_1&=\frac{\sqrt{2}e^{-i\lambda\varphi} \label{eq:vertex-circ-1} e^{i\left(\frac{r-1}{2}\right)\varphi}}{z\sqrt{1-z}} \left(-\lambda k_\perp \left(-\delta_{r \lambda}+(1-z)\delta_{r,-\lambda}\right)\delta_{r^\prime r}+z^2M\delta_{r\lambda}\delta_{r^\prime,-r}\right), \\
	\begin{split}
		V^{r^\prime r\lambda}_2&=\frac{e^{-i\lambda\varphi} e^{i\left(\frac{r-1}{2}\right)\varphi}}{\sqrt{2(1-z)}} \left(\lambda k_\perp\delta_{r^\prime r} + \right.\\
		&\hspace{3cm}\left. + \frac{1}{Mz}\left(\left(M^2z^2-m^2_{\gamma^\prime}\left(1-z\right)\right) \delta_{r\lambda}+k^2_\perp\delta_{r,-\lambda}\right)\delta_{r^\prime,-r}\right),	
	\end{split}
\end{align}
and for the longitudinal dark photon polarization, $\lambda=L$, 
\begin{align}
	V^{r^\prime r L}_1&=\frac{r e^{i\left(\frac{r-1}{2}\right)\varphi}}{z\sqrt{1-z}m_{\gamma^\prime}}\left(k^2_\perp+M^2z^2-m_{\gamma^\prime}^2(1-z)\right)\delta_{r^\prime r}, \\
	V^{r^\prime r L}_2&=\frac{m_{\gamma^\prime} e^{i\left(\frac{r-1}{2}\right)\varphi}}{2z\sqrt{1-z}}\left(rz^2\delta_{r^\prime r}+\frac{k_\perp}{M}(-2+z)\delta_{r^\prime,-r}\right). \label{eq:vertex-long-2}
\end{align}
Note that the vertex function~\eqref{eq:vertex-long-2} contains both terms for the same $\delta_{r^\prime r}$ and for the opposite $\delta_{r^\prime,-r}$ helicities of involved protons.

Evaluating the absolute square of the matrix element~\eqref{eq:Mrpr}, we obtain the sums containing quadratic combinations of vertex functions~\eqref{eq:vertex-func-1}, \eqref{eq:vertex-func-2}
\begin{equation} \label{eq:IJdef}
	\sum_\lambda V^{r^\prime r \lambda}_i\left(V^{r^{\prime\prime} r \lambda}_j\right)^*\equiv \left(I^\prime_{ij}\delta_{r^\prime r}+I^{\prime\prime}_{ij}\delta_{r^\prime,-r}\right)\delta_{r^{\prime\prime}r^\prime}+\left(J^\prime_{ij}\delta_{r^\prime r}+J^{\prime\prime}_{ij}\delta_{r^\prime,-r}\right)\delta_{r^{\prime\prime},-r^\prime},
\end{equation}
where $i,j=\overline{1,2}$ and we introduced the coefficients $I^{\prime(\prime)}_{ij}$, $J^{\prime(\prime)}_{ij}$ that are listed in~\eqref{eq:I11}--\eqref{eq:J22} of appendix~\ref{sec:quadr}. After summing over the common spin index $r$ we obtain the relevant part of the matrix element squared
\begin{equation} \label{eq:sumrl}
	\sum_{r,\lambda} V^{r^\prime r \lambda}_i\left(V^{r^{\prime\prime} r \lambda}_j\right)^* = \left(I^\prime_{ij}+I^{\prime\prime}_{ij}\right)\delta_{r^{\prime\prime}r^\prime}+\left(J^\prime_{ij}+J^{\prime\prime}_{ij}\right)\delta_{r^{\prime\prime},-r^\prime}.
\end{equation}
Then the square of full matrix element~\eqref{eq:Mrpr}
\begin{equation} \label{eq:Msq}
	\sum_{r,\lambda}|\mathcal{M}^{r\lambda}|^2=\left(\frac{\epsilon ez}{H}\right)^2\left(N\sum_{r^\prime}|\mathcal{M}^{r^\prime}|^2	+A\sum_{r^\prime}\mathcal{M}^{r^\prime}\left(\mathcal{M}^*\right)^{-r^\prime}\right)
\end{equation}
contains the normal part, that is proportional to $|\mathcal{M}^{r^\prime}|^2$ and transforms into the cross section of subprocess $p(p-k)p(P_i)\rightarrow X$ in the final answer, and the anomalous spin-flip part, that is proportional to $\mathcal{M}^{r^\prime}\left(\mathcal{M}^*\right)^{-r^\prime}$, with coefficients
\begin{align}
	N\equiv& |F_1|^2\left(I^\prime_{11}+I^{\prime\prime}_{11}\right)+|F_2|^2\left(I^\prime_{22}+I^{\prime\prime}_{22}\right)+\left(F_1F^*_2+F_2F^*_1\right)\left(I^\prime_{12}+I^{\prime\prime}_{12}\right),\\
	A\equiv&
	\left(F_1F^*_2-F_2F^*_1\right)\left(J^\prime_{12}+J^{\prime\prime}_{12}\right).
\end{align}
The anomalous part $A$ does not show up in the calculations for the ``structureless" proton without the magnetic dipole moment. In particular, it is not present in the result of ref.\,\cite{Foroughi-Abari:2021zbm}. It appears for the first time in our work due to account for the proton magnetic dipole moment and its impact on the $pp\gamma^\prime$-vertex in~\eqref{eq:vertex-func-2}. Technically, in the squared matrix element of \eqref{eq:Mrpr} the cross terms proportional to $F_1F^*_2$, $F^*_1F_2$ contain odd number of gamma-matrices. To have non-zero contribution to the trace one needs the even number, and the required term comes from the squared proton wave functions in \eqref{eq:vertex-func-1}, \eqref{eq:vertex-func-2} implying the spin-flip of the scattering proton. 

The anomalous part is an interesting feature of the studied problem but we should keep in mind that our goal is to factorize the final answer and to extract the cross section of inelastic $pp$-scattering subprocess. Comparing the numerical values of Dirac and Pauli electromagnetic form factors according to the fits performed in refs.\,\cite{Faessler:2009tn, Dubnickova:2020heq}, one can observe the following hierarchy, which holds for the largest part of relevant dark photon momentum squared $t=m_{\gamma^\prime}^2$,
\begin{equation} \label{eq:ineq}
	|F_2|^2 > |F_1F^*_2+F_2F^*_1| > |F_1|^2 > |F_1F^*_2-F_2F^*_1|.
\end{equation}
Also we can place an upper limit on the ratio of normal and anomalous quadratic combinations of the vertex functions, 
\begin{equation} \label{eq:upper-lim}
	\frac{|J^\prime_{12}+J^{\prime\prime}_{12}|}{I^\prime_{22}+I^{\prime\prime}_{22}} \simeq \frac{k_\perp}{Mz} < 1\,,
\end{equation}
for all values of $k_\perp$ that give the largest contribution to the cross section. Combining~\eqref{eq:ineq} and~\eqref{eq:upper-lim} we conclude, that the anomalous part is small compared to the normal part and hence can be neglected. It is this \textit{approximation}, that allows us to finish the calculation and present the result in the traditional form.

Finally, along the same lines as it is done in ref.\,\cite{Foroughi-Abari:2021zbm}, we obtain the differential cross section of inelastic bremsstrahlung depending on both Dirac and Pauli form factors as we declared in~\eqref{eq:our-wanted-result} with the master splitting function
\begin{equation} \label{eq:splitfunc}
	w_\text{mas}(z, k^2_\perp)\equiv w_{11}(z, k^2_\perp)|F_1|^2+w_{22}(z, k^2_\perp)|F_2|^2+w_{12}(z, k^2_\perp)\left(F_1F^*_2+F_2F^*_1\right)\,.
\end{equation}
It depends on the three auxiliary splitting functions
\begin{align} 
	w_{11}(z, k^2_\perp)&\equiv \frac{\epsilon^2 \alpha_\text{em}}{2\pi H} \left(z-\frac{z\left(1-z\right)}{H}\left(2M^2+m^2_{\gamma^\prime}\right)+\frac{H}{2zm^2_{\gamma^\prime}}\right),\\
	w_{22}(z, k^2_\perp)&\equiv \frac{\epsilon^2 \alpha_\text{em}}{2\pi H} \frac{m^2_{\gamma^\prime}}{8M^2} \left(z-\frac{z\left(1-z\right)}{H}\left(8M^2+m^2_{\gamma^\prime}\right)+\frac{2H}{zm^2_{\gamma^\prime}}\right),\\ \label{eq:auxsplitfunc}
	w_{12}(z, k^2_\perp)&\equiv \frac{\epsilon^2 \alpha_\text{em}}{2\pi H} \left(\frac{3z}{4}-\frac{3m^2_{\gamma^\prime}z\left(1-z\right)}{2H}\right),
\end{align}
the first of which coincides with $w_\text{FAR}(z, k_\perp^2)$ in~\eqref{eq:FAR-split-func}. 
We should stress that despite neglecting the \textit{new} anomalous part in the square of bremsstrahlung matrix element~\eqref{eq:Msq}, we obtain, as we show later in section~\ref{sec:numerical}, the non-negligible \textit{changes} in the final differential cross section~\eqref{eq:splitfunc}, which are proportional to quadratic combinations $|F_2|^2$ and $\left(F_1F^*_2+F_2F^*_1\right)$. Thus the result still contains interesting modifications, such as the \textit{new} splitting functions $w_{22}(z, k^2_\perp)$ and $w_{12}(z, k^2_\perp)$, that appeared due to taking into account the Pauli electromagnetic form factor.
In section\,\ref{sec:sensitivity} we compare the predicted sensitivities of T2K, DUNE and SHiP to the visible decays of dark photon obtained using the result of this paper~\eqref{eq:our-wanted-result}, \eqref{eq:splitfunc} with the sensitivities based on the Blumlein-Brunner cross section~\eqref{eq:BB-result}, \eqref{eq:BB-split-func} (which is a common choice in experimental papers nowadays).

\section{Numerical results for the full cross sections} \label{sec:numerical}
The full inelastic bremsstrahlung cross section obtained upon integrating~\eqref{eq:our-wanted-result}, \eqref{eq:splitfunc}--\eqref{eq:auxsplitfunc} over the part of $(z, k^2_\perp)\in \left[0;1\right]\times\left[0;1\text{\,GeV}^2\right]$ rectangular, where the inequalities $(k^2_\perp+m^2_{\gamma^\prime})/z^2P^2<0.1$ and $k^2_\perp/(1-z)^2P^2<0.1$ justifying the Taylor expansion of energies $E_k$, $E_{p^\prime}$ hold, can be divided into three terms
\begin{equation} \label{eq:fullcrsec}
	\sigma(pp\rightarrow \gamma^\prime X)=|F_1|^2 \sigma_D + |F_2|^2 \sigma_P + \left(F_1F^*_2+F_2F^*_1\right) \sigma_I.
\end{equation}
Here we introduce the auxiliary Dirac, Pauli and interference cross sections, which are independent of the electromagnetic proton form factor values, 
\begin{equation} \label{eq:auxcrsec}
	\begin{pmatrix}
		\sigma_D \\ \sigma_P \\ \sigma_I
	\end{pmatrix} \equiv \int 
	\begin{pmatrix}
		w_{11}(z, k^2_\perp) \\ w_{22}(z, k^2_\perp) \\ w_{12}(z, k^2_\perp)
	\end{pmatrix} 
	F^2_\text{virt}(z, k^2_\perp)\sigma_\text{NSD}(\bar s) H_0(z, k^2_\perp) H_1(z, k^2_\perp)\dd z \dd k^2_\perp. 
\end{equation}
Here the functions which limit the region of integration,
\begin{align}
	H_0(z, k^2_\perp)&\equiv \Theta\left(0.1-\frac{k^2_\perp+m^2_{\gamma^\prime}}{z^2P^2}\right),\\
	H_1(z, k^2_\perp)&\equiv \Theta\left(0.1-\frac{k^2_\perp}{(1-z)^2P^2}\right)
\end{align}
are constructed from the Heaviside step-function $\Theta(x)$, that is unity for positive $x$ and null for negative $x$. Throughout this section for simplicity we fix the kinetic mixing parameter $\epsilon=1$. For realistic estimates, the inelastic bremsstrahlung cross sections values must be multiplied by factor $\epsilon^2$.

We put below the numerical fits for the auxiliary cross sections in $\mu\text{b}$ depending on the parameter $y\equiv m_{\gamma^\prime}/\text{GeV}$ for the fixed value of the incident proton momentum $P=120\,\text{GeV}$
\begin{align}
	\sigma_D/\mu\text{b}&=-0.0449/y^3 + 19.5/y^2 - 28.7/y + 24.8 - 10.0 y + 1.51 y^2,\label{eq:fit1}\\
	\sigma_P/\mu\text{b}&=0.0192/y^3 - 0.421/y^2 + 3.20/y + 0.392 - 0.885 y + 0.260 y^2,\\ 
	\sigma_I/\mu\text{b}&=0.0209/y^3 - 0.467/y^2 + 3.66/y - 1.10 - 0.548 y + 0.221 y^2. \label{eq:fit3}
\end{align}
The numerical approximation~\eqref{eq:fit1}--\eqref{eq:fit3} works well in the region $0.1<y<2.3$, although the process of inelastic bremsstrahlung dominates dark photon production only for $0.4<y<1.8$. More complicated numerical fits, allowing to vary both  the dark photon mass $m_{\gamma^\prime}$ and the incident proton momentum $P$ and their relative deviations can be found in appendix~\ref{sec:fits}.
Using~\eqref{eq:fit1}--\eqref{eq:fit3} or \eqref{eq:fit-small-D}--\eqref{eq:fit-large-I} the full inelastic cross section~\eqref{eq:fullcrsec} can be obtained without integration only by choosing the appropriate fit for the Dirac and Pauli electromagnetic form factors. 
 
\begin{figure}[t]
	\begin{center}
		\includegraphics[width=0.7\textwidth]{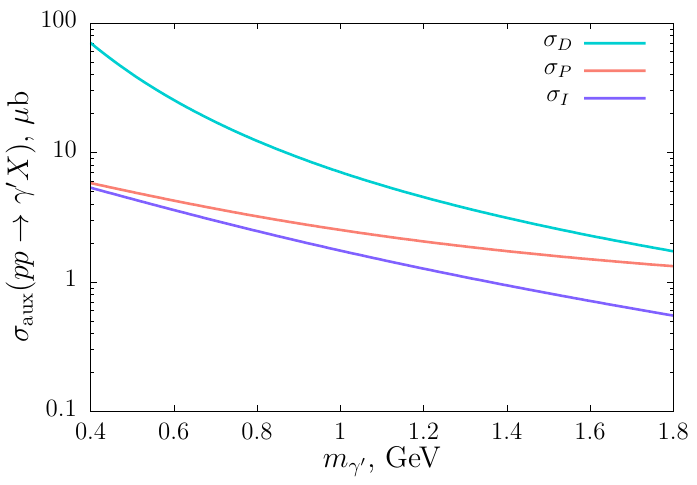}
		\caption{The auxiliary Dirac (turquoise line), Pauli (salmon line) and interference (violet line) cross sections~\eqref{eq:auxcrsec} as functions of dark photon mass $m_{\gamma^\prime}$ for the incident proton momentum $P=120\,\text{GeV}$.} 
		\label{fig:auxcrsec}
	\end{center}
\end{figure}

Figure\,\ref{fig:auxcrsec} 
shows the dependence of auxiliary cross sections~$\sigma_D$, $\sigma_P$ and $\sigma_I$ on dark photon mass $m_{\gamma^\prime}$ and allows us to compare their relative contributions to the full inelastic bremsstrahlung cross section for the naive choice of both Dirac and Pauli form factors equal to unity.

It can be seen that, if one forgets about the form factor  values, the Dirac contribution $\sigma_D$ is indeed greater than the other two. So, naively, the Pauli $\sigma_P$ and the interference $\sigma_I$ auxiliary cross sections, obtained in this work, could be neglected in comparison with $\sigma_D$. 

\begin{figure}[t]
	\begin{center}
		\includegraphics[width=0.7\textwidth]{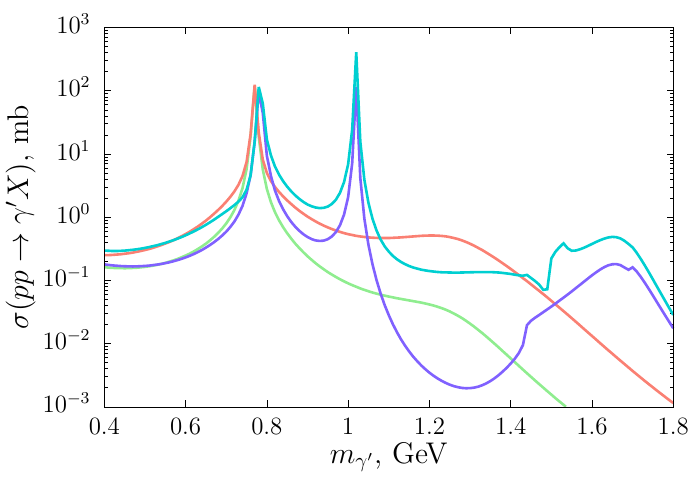}
		\caption{Full inelastic bremsstrahlung cross section as the function of dark photon mass $m_{\gamma^\prime}$. The cross section calculated using only the Dirac form factor~\eqref{eq:FAR-result} and fit~\cite{Faessler:2009tn} is depicted with green line, the same calculation using fit~\cite{Dubnickova:2020heq} is shown in violet. Salmon line presents the result of this paper~\eqref{eq:our-wanted-result} including both Dirac and Pauli form factors from the conservative fit~\cite{Faessler:2009tn}, turquoise line shows the same result, but plotted using the fit from ref.\,\cite{Dubnickova:2020heq}. The incident proton momentum is fixed at $P=120\,\text{GeV}$.} 
		\label{fig:fullcrseccomp}
	\end{center}
\end{figure}

Of course, such approach is incomplete, so we restore the values of Dirac $F_1(m^2_{\gamma^\prime})$ and Pauli $F_2(m^2_{\gamma^\prime})$ electromagnetic form factors and present the comparison of the full inelastic bremsstrahlung cross sections computed using fits~\cite{Faessler:2009tn,Dubnickova:2020heq} in figure\,\ref{fig:fullcrseccomp}.
It is evident from figure~\ref{fig:fullcrseccomp} that for any choice of the fit for electromagnetic proton form factors~\cite{Faessler:2009tn} or~\cite{Dubnickova:2020heq}, the full inelastic bremsstrahlung cross section~\eqref{eq:our-wanted-result} with the master splitting function~\eqref{eq:splitfunc} incorporating both Dirac and Pauli form factors is noticeably larger than the cross section~\eqref{eq:FAR-result} obtained earlier in~\cite{Foroughi-Abari:2021zbm} with only the Dirac form factor. Moreover, due to larger set of vector meson resonances in the basis of fit~\cite{Dubnickova:2020heq} as compared to~\cite{Faessler:2009tn}, the former case exhibits a larger region in dark photon mass where the full cross section is amplified due to the resonant behavior of the electromagnetic form factors.

\section{Experimental sensitivity of T2K, DUNE and SHiP} \label{sec:sensitivity}
In this section we estimate the expected sensitivity to visible dark photon decays of the off-axis near detector ND280 of T2K~\cite{Araki:2023xgb}, the Multi-Purpose Detector (MPD) of DUNE~\cite{Berryman:2019dme} and the decay volume of SHiP~\cite{SHiP:2020vbd}. In general, for all three experiments we follow the procedures described in the appendix A of~\cite{Araki:2023xgb} and the appendix B of~\cite{Gorbunov:2023jnx} using the parameters listed in table~\ref{tab:exp-params}.
\begin{table}[t]
	\centering
		\begin{tabular}{|c|c|c|c|c|}
			\hline
			\multicolumn{2}{|c|}{ } & \textbf{T2K} & \textbf{DUNE} & \textbf{SHiP} \\
			\hline
			\multicolumn{2}{|c|}{$P$,\,GeV} & 30 & 120 & 400 \\
			\hline
			\multicolumn{2}{|c|}{Detector shape} & \makecell{rectangular at\\$\Theta_{ND}=2^\circ$ away from\\ beam direction, \\fig.~8 in~\cite{Araki:2023xgb}} & \makecell{cylinder with axis\\ perpendicular to\\beam  direction,\\fig.~2.1 in~\cite{Berryman:2019dme}} & \makecell{truncated pyramid, \\figs.~1, 4 in~\cite{Miano:2020gjq}} \\
			\hline
			\multirow{3}{*}{\makecell{Detector\\geometry}} & $w$,\,m & 2.4 & 5 & min 1.5, max 5 \\
			\hhline{~----}
			& $h$,\, m & 2.4 & \multirow{2}{*}{5\,(diameter)} & min 4.3, max 11 \\
			\hhline{~--~-}
			& $d$,\, m & 5.8 & & 50 \\
			\hline
			\multicolumn{2}{|c|}{$L_\text{min}$,\,m} & 280.1 & 579 & 60 \\
			\hline
			\multicolumn{2}{|c|}{$N_\text{POT}$} & $3.7\cdot10^{22}$ & $1.47\cdot10^{22}$ & $2\cdot10^{20}$ \\
			\hline
			\multicolumn{2}{|c|}{Target material} & \multicolumn{2}{c|}{C (graphite)} & Mo \\
			\hline
			\multicolumn{2}{|c|}{Comments} & \makecell{efficiency\\ $0.25\cdot0.82$} & \makecell{in contrast to~\cite{Berryman:2019dme}\\ we numerically\\ integrate over the\\ actual position\\ of cylinder} & \makecell{cut in dark photon\\ momentum space\\ $k^2_\perp+z^2P^2+m^2_{\gamma^\prime}>$\\$>(2\text{\,GeV})^2$} \\
			\hline
		\end{tabular}    
	\caption{The detector characteristics of T2K, DUNE and SHiP experiments that were used for estimating their sensitivity to visible dark photon decays.}
	\label{tab:exp-params}
\end{table}
We will study the sensitivity only in the region of dark photon mass $m_{\gamma^\prime}$ where the inelastic bremsstrahlung is the dominant production channel (0.4--1.8\,GeV), so we neglect the inputs of two other production channels, such as meson decays and QCD Drell-Yan process, in this region. This rough estimate is enough to reach our final goal, e.g. to demonstrate the significant changes in the expected sensitivities obtained using the result of this work~\eqref{eq:our-wanted-result}, \eqref{eq:splitfunc}--\eqref{eq:auxsplitfunc} in contrast to the widespread approach of Blumlein and Brunner~\eqref{eq:BB-result}, \eqref{eq:BB-split-func}.

Table~\ref{tab:exp-params} contains the geometric parameters of T2K, DUNE and SHiP detectors, such as shape, width $w$ (along horizontal $x$-axis), height $h$ (along vertical $y$-axis), depth $d$ (along $z$-axis aligned with the beam direction) and the distance along the $z$-axis from the interaction point to the nearest detector side $L_\text{min}$. For T2K we reconstruct the rectangular detector position and geometry as described in~\cite{Araki:2023xgb}. For DUNE in contrast to~\cite{Berryman:2019dme} we calculate the number of events for the actual position of the cylindrical detector with the cylinder axis being perpendicular to the beam direction. This gives the MPD constant width $w$ and variable height above the beam line as a function of the distance $z_\text{det}$ from the detector front in $z$-direction, 
\begin{equation} \label{eq:h-dune}
	h(z_\text{det})=2\sqrt{z_\text{det}\left(d-z_\text{det}\right)}.
\end{equation}
Since SHiP decay vessel has the form of truncated pyramid, in this case the height and the width grow linearly from the minimum to the maximum values listed in table~\ref{tab:exp-params}.

The radius of circumference that reach the dark photons of a given momentum at position $z_\text{det}$ is 
\begin{equation} \label{eq:radius}
	r(z, k_\perp, z_\text{det}) = \frac{k_\perp}{zP}(L_\text{min}+z_\text{det}).
\end{equation}
The probability of dark photon detection depends on the ratio $P_\theta(z, k^2_\perp, z_\text{det})$ of dark photons with fixed 3-momentum and position $z_\text{det}$ inside the detector. In the case of T2K we adopt $P_\theta(z, k^2_\perp, z_\text{det})$ presented in~\cite{Araki:2023xgb}. The expression $P_\theta(z, k^2_\perp, z_\text{det})$ for SHiP summarizes the possible ratios of dark photon circumference length with radius $r(z, k_\perp, z_\text{det})$~\eqref{eq:radius} that lies inside the rectangular slice of SHiP decay vessel with sides $w_d\equiv w(z_\text{det})$, $h_d\equiv h(z_\text{det})$ and diagonal $\rho_d\equiv \sqrt{w_d^2+h_d^2}$ at position $z_\text{det}$ on the $z$-axis from its beginning, 
\begin{equation} \label{eq:Ptheta}
	P_\theta(z, k^2_\perp, z_\text{det})=
	\begin{cases}
		1, & 2r(z, k_\perp, z_\text{det}) \leq w_d, \\
		1-A(w_d), & w_d < 2r(z, k_\perp, z_\text{det}) \leq h_d, \\
		1-A(w_d)-A(h_d), & h_d < 2r(z, k_\perp, z_\text{det}) \leq \rho_d, \\
		0, & \rho_d < 2r(z, k_\perp, z_\text{det}),
	\end{cases}
\end{equation}
where
\begin{equation}
	A(l)\equiv \frac{2}{\pi}\arccos\left(\frac{l}{2r(z, k_\perp, z_\text{det})}\right).
\end{equation}
Unlike SHiP decay vessel, the width of DUNE MPD is greater than its height for all $z_\text{det}$, so in this case one first needs to exchange in~\eqref{eq:Ptheta} $w_d\leftrightarrow h_d$ and then use constant $w_d=w$ and variable $h_d$, defined in~\eqref{eq:h-dune}.

Mixing with the SM photon allows the dark photon to decay into pairs of charged leptons of mass $m_l$ or into hadrons with the decay widths~\cite{Miller:2021ycl}
\begin{align} \label{eq:decwid}
	\Gamma_\text{lep}(m_l) &= \frac{\alpha \epsilon^2}{3} m_{\gamma^\prime} \sqrt{1-\frac{4m_l^2}{m_{\gamma^\prime}^2}}\left(1+\frac{2m_l^2}{m^2_{\gamma^\prime}}\right),\\
	\Gamma_\text{had} &= \Gamma_\text{lep}(m_\mu) R(m_{\gamma^\prime}),
\end{align}
where the R-ratio
\begin{equation}
	R(\sqrt{s})\equiv\frac{\sigma(e^+e^-\rightarrow \text{hadrons})}{\sigma(e^+e^-\rightarrow \mu^+\mu^-)}
\end{equation}
is obtained from experimental data~\cite{ParticleDataGroup:2022pth}. The total decay width is then given by (we ignore possible decays to invisible channels, e.g. to lighter particles from the dark sector)
\begin{equation}
	\Gamma_\text{tot}=\Gamma_\text{lep}(m_e)+\Gamma_\text{lep}(m_\mu)+\Gamma_\text{had}.
\end{equation}
The dark photon decay length projected onto axis, directed along the incident proton beam, reads
\begin{equation}
	L(z)=\frac{zP}{\Gamma_\text{tot} m_{\gamma^\prime}},
\end{equation}
Thus the probability of dark photon to decay inside the detector can be obtained by numerical integration over slices of the detector volume in $z$-direction,  
\begin{equation} \label{eq:Pdet}
	P_\text{det}(z, k^2_\perp) = \frac{e^{-\frac{L_\text{min}}{L(z)}}}{L(z)} \int_{0}^{d} \dd{z_\text{det}} e^{-\frac{z_\text{det}}{L(z)}} P_\theta(z, k^2_\perp, z_\text{det}).
\end{equation}
In the case of SHiP we additionally multiply the right-hand-side by the factor representing the cut on dark photon momenta $\Theta(k^2_\perp+z^2P^2+m^2_{\gamma^\prime}-(2\text{\,GeV})^2)$.

Following~\cite{SHiP:2020vbd}, for each experiment we estimate the luminosity as
\begin{equation} \label{eq:lumin}
	\mathcal{L}\equiv\frac{N_\text{POT}}{\sigma_\text{inel}},
\end{equation}
where $N_\text{POT}$ is the number of protons on target listed in table~\ref{tab:exp-params} and the inelastic proton-nucleus cross section~\cite{SHiP:2015vad}
\begin{equation}
	\sigma_\text{inel}\equiv \frac{A}{\lambda_\text{int}\rho N_A},
\end{equation}
is constructed from the tabular values of atomic mass $A$ in g\,mol$^{-1}$, nuclear interaction length $\lambda_\text{int}$ in cm, material density $\rho$ in g\,cm$^{-3}$ which are given in~\cite{PDG:Materials}. We adopt the following values for $\sigma_\text{inel}$: 10.7\,mb (SHiP, Mo target) and 19.3\,mb (T2K, DUNE, graphite target). We note here that the total $pp$-scattering cross section~\cite{Denisov:1971jb,Carroll:1975xf}, as well as the absorption cross section of proton on graphite nuclei~\cite{Denisov:1973zv,Carroll:1978hc} feebly depend on the incident proton energy for the energy range from T2K to DUNE beam energies. This is why we take the same value of inelastic proton-nucleus cross section for these experiments.

To compute the number of events, it is more convenient to integrate over the dark photon momentum polar angle $\theta$,
\begin{equation}
	\tan \theta\equiv\frac{k_\perp}{zP},
\end{equation}
rather than over the transverse dark photon momentum squared $k^2_\perp$. Thus for every experiment we roughly estimate the lower and upper limits of integration over polar angle as $\theta_l$ and $\theta_u$. There is no need to do it exactly, since the real borders of the detector are already taken into account via the detection probability $P_\text{det}(z, k^2_\perp)$~\eqref{eq:Pdet}. The Jacobian for this change of variables reads 
\begin{equation} \label{eq:Jacob}
	J(z, \theta)\equiv 2z^2P^2 \tan \theta \left(1+\tan^2 \theta\right).
\end{equation}

Combining~\eqref{eq:Pdet}, \eqref{eq:lumin} and \eqref{eq:Jacob}, allows us to estimate the number of signal events as 
\begin{equation}
	N_\text{ev} = \mathcal{L} \int_{0}^{1}\dd z \int_{\theta_l}^{\theta_u}\dd \theta \left(\mathcal{F}^2\cdot \frac{\dd^2 \sigma_0}{\dd z \dd k^2_\perp}\right) J(z, \theta) P_\text{det}(z, k^2_\perp(z, \theta)),
\end{equation}
where the parentheses with dot denotes the scalar product of the vector, made of electromagnetic form factor quadratic combinations
\begin{equation}
	\mathcal{F}^2\equiv\{|F_1|^2, |F_2|^2, F_1F^*_2+F_2F^*_1\},
\end{equation}
and the second derivative of the vector, formed by the auxiliary cross sections~\eqref{eq:auxcrsec},
\begin{equation}
	\sigma_0\equiv\{\sigma_D, \sigma_P, \sigma_I\}.
\end{equation}
Additionally, in the case of T2K we multiply the overall number of events by the factor $0.25\cdot0.82$, accounting for the effective volume of the detector and its efficiency~\cite{Araki:2023xgb}.

\begin{figure}[t]
	\begin{center}
		\includegraphics[width=0.7\textwidth]{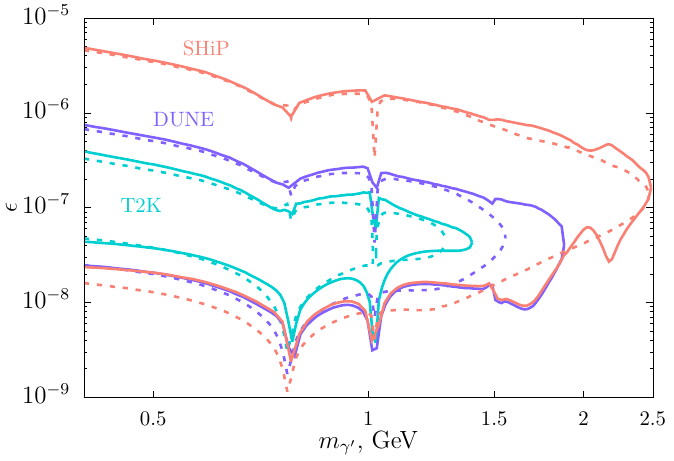}
		\caption{The expected sensitivity of T2K, DUNE and SHiP experiments to visible dark photon decays. Solid lines correspond to the result of this work~\eqref{eq:our-wanted-result}, \eqref{eq:splitfunc} with Dirac and Pauli electromagnetic form factors~\cite{Dubnickova:2020heq}. Dashed lines show the widely accepted sensitivity estimates based on the Blumlein and Brunner approximation~\eqref{eq:BB-result}--\eqref{eq:BB-split-func} with the fit for Dirac form factor~\cite{Faessler:2009tn}.} 
		\label{fig:sensitivity}
	\end{center}
\end{figure}

In figure~\ref{fig:sensitivity}
we show the expected curves of T2K, DUNE and SHiP sensitivities to visible dark photon decays that correspond (with zero background assumption) to 3 signal events. If dark photons are not observed, the regions inside these curves will be excluded at the 95\% CL. For each experiment we compare the sensitivity based on the result of this work~\eqref{eq:our-wanted-result}, \eqref{eq:splitfunc} using both electromagnetic form factors from~\cite{Dubnickova:2020heq} (solid line) with the previous estimates of sensitivity provided in~\cite{Araki:2023xgb,Berryman:2019dme,SHiP:2020vbd} (dashed line), where the cross sections were calculated within the framework of Blumlein and Brunner~\eqref{eq:BB-result}--\eqref{eq:BB-split-func} and using only Dirac form factor fit~\cite{Faessler:2009tn}. It can be seen that including the terms with Pauli form factor significantly extends the region available for dark photon searches and changes the sensitivity to the kinetic mixing parameter $\epsilon$ at a given dark photon mass $m_{\gamma^\prime}$.

\section{Discussion} \label{sec:discuss}
To summarize, we have studied the new contributions to inelastic bremsstrahlung cross section associated with the Pauli form factor term in the $pp\gamma^\prime$ vertex. Working within the approach formulated in~\cite{Foroughi-Abari:2021zbm}, we have obtained two new auxiliary splitting functions. We have provided the numerical fits for integrated auxiliary cross sections as functions of incident proton momentum $P$ and dark photon mass $m_{\gamma^\prime}$. They must be multiplied by the form factor products and summed up to obtain the total production cross section. Using two different sets of proton Dirac and Pauli electromagnetic form factors~\cite{Faessler:2009tn,Dubnickova:2020heq}, we have shown that the impact of new terms is significant regardless of the chosen fit. We have also updated the estimates for experimental sensitivity of T2K, DUNE and SHiP to visible dark photon decays and demonstrated that our results considerably change the regions in dark photon mass $m_{\gamma^\prime}$ which can be explored in these experiments.  

It is worth noting that the final result for the full cross section of inelastic bremsstrahlung and for experimental sensitivity to dark photon depends on the utilized fit for the electromagnetic Dirac and Pauli form factors. The main reason is that in the unphysical region these form factors are obtained not from the direct measurements, but rather from the interpolation of measured form factors in spacelike and timelike regions. Nevertheless, to obtain the robust predictions for the new physics signal, it is very important to reduce the existing uncertainties in the electromagnetic form factors values, e.g. by exploring them in the unphysical region of proton pair momentum squared below the proton-antiproton threshold. This study may be performed in $p\bar p$ collisions in PANDA\,\cite{PANDA:2021ozp} at FAIR with production of $e^+e^-$ pairs associated with pions\,\cite{Kuraev:2012ca,Adamuscin:2006dfq}. Thus the studies of hypothetical dark photon can encourage the investigation of electromagnetic form factors in the unphysical region within the framework of SM.

The result obtained in this work implies changes of the predicted sensitivity not only for T2K, DUNE and SHiP, but also for other experiments, in which proton bremsstrahlung makes significant contribution to dark photon production. We find it important to update the existing sensitivity curves for such experiments searching for visible decays of dark photons, summarized e.g. in~\cite{Ovchynnikov:2023cry}. Another application for our refined calculation is the search for heavy millicharged particles from the atmosphere, recently described in~\cite{Du:2022hms,Wu:2024iqm} and perfomed with the use of Blumlein and Brunner result~\eqref{eq:BB-result}--\eqref{eq:BB-split-func}.

As the further development of this work, it might be also important to estimate the contribution of virtual $\Delta^+$-resonance to the inelastic proton bremsstrahlung. Analogous correction to the two-photon exchange in electron-proton scattering was considered in~\cite{Kondratyuk:2005kk}, but turned out to be relatively small. In the case of inelastic proton bremsstrahlung it can be potentially enhanced for nearly on-shell $\Delta^+$.

Before submitting the text of this paper to arXiv, we became aware of a recent preprint~\cite{Foroughi-Abari:2024xlj} also considering dark photon bremsstrahlung. Its analysis partly overlaps with our results of the section~\ref{sec:new-factorization} which are based on talks~\cite{Mosphys,SessiyaOFNRAN} and conference paper~\cite{Gorbunov:2024}. The latter contained the idea of the calculation from section~\ref{sec:new-factorization} and sketched the results for auxiliary splitting functions~\eqref{eq:splitfunc}--\eqref{eq:auxsplitfunc}, though their detailed derivation is published here for the first time.

\acknowledgments

We are indebted to Y.~H.~Lin, S.~Dubnicka and A.-Z.~Dubnickova who shared with us the files with tabulated electromagnetic proton form factors. We are grateful to A.~Arbuzov, V.~Barinov, S.~Demidov, S.~Godunov, S.~Kulagin, E.~Kuzminskii, R.~Lee, M.~Matveev, O.~Teryaev, V.~Troitsky and M.~Vysotsky for valuable discussions and interesting comments. This work is supported in the framework of the State project ``Science'' by the Ministry of Science and Higher Education of the Russian Federation under the contract 075-15-2024-541. EK thanks the Foundation for the Advancement of Theoretical Physics and Mathematics “BASIS” for the PhD fellowship No. 21-2-10-37-1.
\appendix
\section{Quadratic combinations of vertex functions} \label{sec:quadr}
Using the explicit form of vertex functions~\eqref{eq:vertex-circ-1}--\eqref{eq:vertex-long-2}, we obtain the following set of quadratic combinations defined in Eq.\,\eqref{eq:IJdef}:
\begin{align}
	I^\prime_{11} &= \frac{\left(k^2_\perp+M^2z^2-m^2_{\gamma^\prime}\left(1-z\right)\right)^2+2k^2_\perp m^2_{\gamma^\prime}\left(1+\left(1-z\right)^2\right)}{m^2_{\gamma^\prime}z^2\left(1-z\right)}, \label{eq:I11} \\
	I^{\prime\prime}_{11} &= \frac{2M^2z^2}{1-z}, \\
	J^\prime_{11} &= -J^{\prime\prime}_{11} = \frac{2r^\prime k_\perp M}{1-z}, \\
	I^\prime_{12} &= I^\prime_{21} = \frac{3k^2_\perp+M^2z^2-m^2_{\gamma^\prime}\left(1-z\right)}{2\left(1-z\right)}, \\
	I^{\prime\prime}_{12} &= I^{\prime\prime}_{21} = \frac{M^2z^2}{1-z}-m^2_{\gamma^\prime}, \\
	J^\prime_{12} &= -J^{\prime\prime}_{21} = -\frac{r^\prime k_\perp \left(k^2_\perp-M^2z^2+m^2_{\gamma^\prime}\left(1-z\right)\right)}{2Mz\left(1-z\right)}, \\
	J^{\prime\prime}_{12} &= - J^{\prime}_{21} = -\frac{r^\prime k_\perp Mz}{1-z}, \\
	I^\prime_{22} &= \frac{4k^2_\perp+m^2_{\gamma^\prime}z^2}{4\left(1-z\right)}, \\
	I^{\prime\prime}_{22} &= \frac{2\left(k^2_\perp-M^2z^2+m^2_{\gamma^\prime}\left(1-z\right)\right)^2+k^2_\perp z^2\left(4M^2+m^2_{\gamma^\prime}\right)}{4M^2z^2\left(1-z\right)}, \\
	J^\prime_{22} &= - J^{\prime\prime}_{22} = -\frac{r^\prime k_\perp \left(2k^2_\perp-2M^2z^2-m^2_{\gamma^\prime}\left(z^2-2\right)\right)}{4Mz\left(1-z\right)}. \label{eq:J22}
\end{align}

\section{Numerical fits for auxiliary cross sections} \label{sec:fits}
Below are numerical fits for the auxiliary cross sections~\eqref{eq:auxcrsec} in $\mu\text{b}$ as functions of the parameters $x\equiv P/\text{GeV}$ and $y\equiv m_{\gamma^\prime}/\text{GeV}$ obtained separately for small incident proton momenta, $x\in[15,50]$,
\begin{align}
\begin{split}
    \sigma_D/\mu\text{b} = D_{-3}/y^3+&D_{-2}/y^2+\hat{D}_{-2} x^{1/4}/y^2 + D_{-1}/y + \hat{D}_{-1} x^{1/4}/y + \label{eq:fit-small-D}\\
    +&D_0+\hat{D}_0 x^{1/4} + D_1y + \hat{D}_1 x^{1/4} y + D_2 y^2, 
    \end{split}
    \\
    \sigma_P/\mu\text{b} = P_{-3}/y^3 +& P_{-2}/y^2 + P_{-1}/y + P_0 + \hat{P}_0 x^{1/4} + P_1 y + \hat{P}_1 x^{1/4}y + P_2 y^2, 
    \label{eq:fit-small-P}
    \\
    \begin{split}
    \sigma_I/\mu\text{b}= \mathcal{I}_{-3}/y^3 +& \mathcal{I}_{-2}/y^2 + \mathcal{I}_{-1}/y + \hat{\mathcal{I}}_{-1} y^{-1}\ln{x}+\\ +& \mathcal{I}_0 + \hat{\mathcal{I}}_0 \ln{x} +\mathcal{I}_1 y + \hat{\mathcal{I}}_1 y \ln{x} + \mathcal{I}_2 y^2, \label{eq:fit-small-I}
\end{split}
\end{align}
and for large incident proton momenta, $x\in[50,450]$,
\begin{align}
\begin{split}
    \sigma_D/\mu\text{b} = D_{-3}/y^3 +& D_{-2}/y^2 + \hat{D}_{-2}x^{1/4}/y^2 + D_{-1}/y + \hat{D}_{-1}x^{1/4}/y + \label{eq:fit-large-D}\\ +& D_0 + \hat{D}_0 x^{1/4} + D_1 y + D_2 y^2,
    \end{split}
    \\
    \sigma_P/\mu\text{b} = P_{-3}/y^3 +& P_{-2}/y^2+P_{-1}/y+P_0+\hat{P}_0x^{1/4}+P_1y+P_2y^2,
    \label{eq:fit-large-P}
    \\
    \begin{split}
    \sigma_I/\mu\text{b} = \mathcal{I}_{-3}/y^3+&\mathcal{I}_{-2}/y^2+\hat{\mathcal{I}}_{-2}x^{1/4}/y^2+ \mathcal{I}_{-1}/y + \hat{\mathcal{I}}_{-1}x^{1/4}/y + \\+&\mathcal{I}_0 + \hat{\mathcal{I}}_0 x^{1/4} + \mathcal{I}_1y + \mathcal{I}_2y^2, \label{eq:fit-large-I}
    \end{split}
\end{align}
where all parameter values are listed in tables~\ref{tab:fit-small-P} and~\ref{tab:fit-large-P} correspondingly.
\begin{table}[t]
    \centering
    \begin{tabular}{|c|c||c|c||c|c||c|c|}
        \hline
        $D_{-3}$ & $-6.20\cdot 10^{-2}$ & $D_1$ & $-4.93$ & $\hat{P}_0$ & 1.98 & $\hat{\mathcal{I}}_{-1}$ & $5.01\cdot 10^{-2}$\\
        \hline
        $D_{-2}$ & 4.48 & $\hat{D}_1$ & $-1.22$ & $P_1$ & 0.470 & $\mathcal{I}_0$ & $-4.59$\\
        \hline
        $\hat{D}_{-2}$ & 5.42 & $D_2$ & 1.12 & $\hat{P}_1$ & $-0.627$ & $\hat{\mathcal{I}}_0$ & 0.875\\
        \hline
        $D_{-1}$ & $-9.53$ & $P_{-3}$ & $1.72\cdot 10^{-2}$ & $P_2$ & 0.261 & $\mathcal{I}_1$ & 0.891\\
        \hline
        $\hat{D}_{-1}$ & $-6.60$ & $P_{-2}$ & $-0.376$ & $\mathcal{I}_{-3}$ & $1.98\cdot 10^{-2}$ & $\hat{\mathcal{I}}_1$ & $-0.348$\\
        \hline
        $D_0$ & 7.59 & $P_{-1}$ & 2.83 & $\mathcal{I}_{-2}$ & $-0.441$ & $\mathcal{I}_2$ & 0.183\\
        \hline
        $\hat{D}_0$ & 5.35 & $P_0$ & $-4.29$ & $\mathcal{I}_{-1}$ & 3.28 & & \\
        \hline
    \end{tabular}
    \caption{Fitted parameter values for auxiliary cross sections~\eqref{eq:fit-small-D}--\eqref{eq:fit-small-I} and incident proton momentum $15\text{ GeV}<P<50\text{ GeV}$.}
    \label{tab:fit-small-P}
\end{table}
\begin{table}[t]
    \centering
    \begin{tabular}{|c|c||c|c||c|c||c|c|}
        \hline
        $D_{-3}$ & $-4.52\cdot 10^{-2}$ & $D_1$ & $-10.2$ & $P_1$ & $-0.902$ & $\mathcal{I}_0$ & $-1.32$\\
        \hline
        $D_{-2}$ & 15.3 & $D_2$ & 1.53 & $P_2$ & 0.276 & $\hat{\mathcal{I}}_0$ & $6.41\cdot 10^{-2}$\\
        \hline
        $\hat{D}_{-2}$ & 1.24 & $P_{-3}$ & $1.99\cdot 10^{-2}$ & $\mathcal{I}_{-3}$ & $2.16\cdot 10^{-2}$ & $\mathcal{I}_1$ & $-0.565$\\
        \hline
        $D_{-1}$ & $-24.5$ & $P_{-2}$ & $-0.435$ & $\mathcal{I}_{-2}$ & $-0.432$ & $\mathcal{I}_2$ & 0.226\\
        \hline
        $\hat{D}_{-1}$ & $-1.28$ & $P_{-1}$ & 3.30 & $\hat{\mathcal{I}}_{-2}$ & $-1.28\cdot 10^{-2}$ & & \\
        \hline
        $D_0$ & 22.9 & $P_0$ & $-0.694$ &$\mathcal{I}_{-1}$ & 3.08 & & \\
        \hline
        $\hat{D}_0$ & 0.664 & $\hat{P}_0$ & 0.289 & $\hat{\mathcal{I}}_{-1}$ & 0.182 & & \\
        \hline
    \end{tabular}
    \caption{Fitted parameter values for auxiliary cross sections\,\eqref{eq:fit-large-D}--\eqref{eq:fit-large-I} and incident proton momentum $50\text{ GeV}<P<450\text{ GeV}$.}
    \label{tab:fit-large-P}
\end{table}

 We also present the contour plots with the relative deviation of the direct result of integration from the numerical fits for small and large incident proton momenta in figures
\begin{figure}[t]
	\begin{center}
		\begin{subfigure}{0.32\textwidth}
			\centering
			\includegraphics[width=\textwidth]{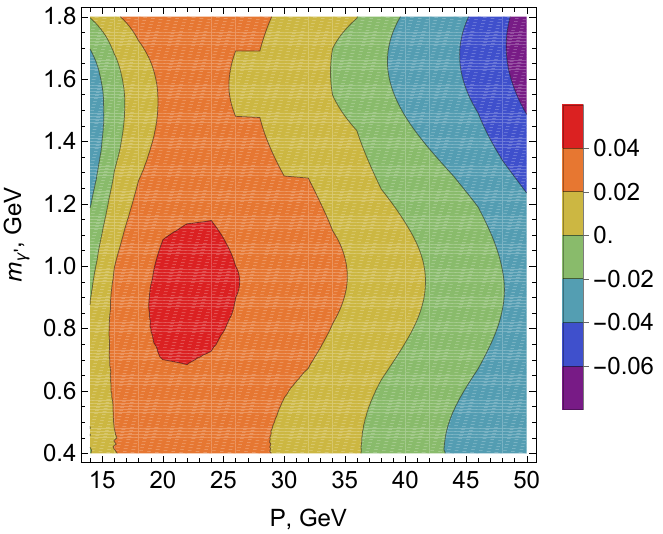}
			\caption{}
			\label{fig:rel-dev-D-small}
		\end{subfigure}
		\hfill
		\begin{subfigure}{0.32\textwidth}
			\centering
			\includegraphics[width=\textwidth]{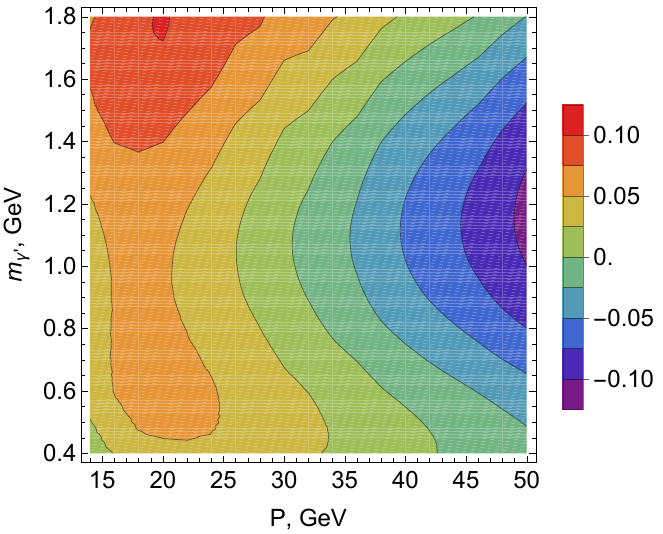}
			\caption{}
			\label{fig:rel-dev-P-small}
		\end{subfigure}
        \hfill
        \begin{subfigure}{0.32\textwidth}
			\centering
			\includegraphics[width=\textwidth]{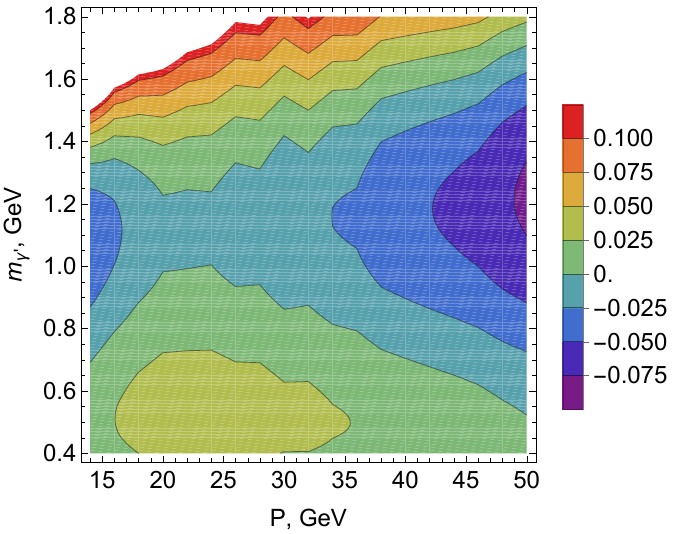}
			\caption{}
			\label{fig:rel-dev-I-small}
		\end{subfigure}
	\end{center}
	\caption{Relative deviation of the integrated (a) Dirac, (b) Pauli, (c) interference auxiliary cross sections~\eqref{eq:auxcrsec} from numerical fits~\eqref{eq:fit-small-D}--\eqref{eq:fit-small-I} for incident proton momentum $15\text{ GeV}<P<50\text{ GeV}$.}
        \label{fig:rel-dev-small}
\end{figure}
\begin{figure}[t]
	\begin{center}
		\begin{subfigure}{0.32\textwidth}
			\centering
			\includegraphics[width=\textwidth]{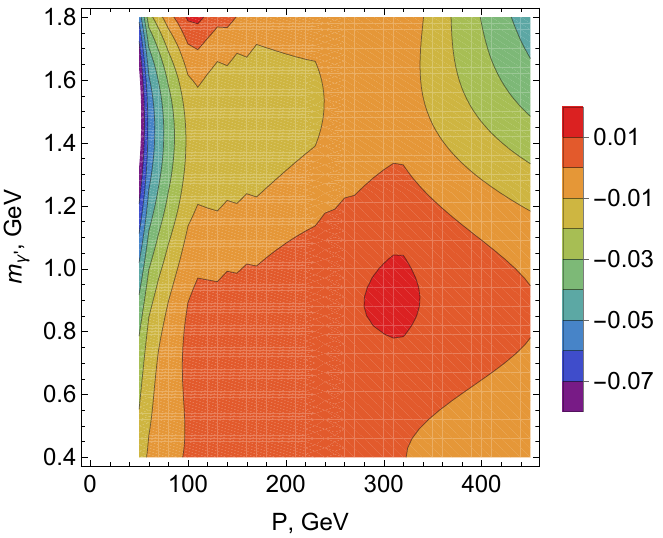}
			\caption{}
			\label{fig:rel-dev-D-large}
		\end{subfigure}
		\hfill
		\begin{subfigure}{0.32\textwidth}
			\centering
			\includegraphics[width=\textwidth]{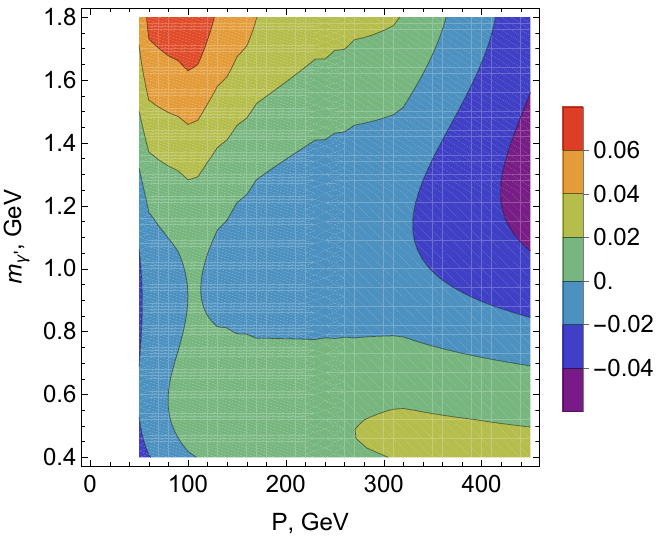}
			\caption{}
			\label{fig:rel-dev-P-large}
		\end{subfigure}
        \hfill
        \begin{subfigure}{0.32\textwidth}
			\centering
			\includegraphics[width=\textwidth]{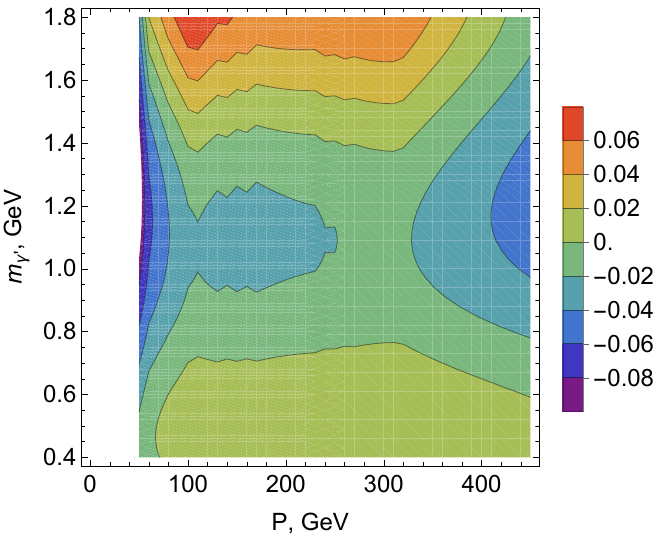}
			\caption{}
			\label{fig:rel-dev-I-large}
		\end{subfigure}
	\end{center}
	\caption{Relative deviation of the integrated (a) Dirac, (b) Pauli, (c) interference auxiliary cross sections~\eqref{eq:auxcrsec} from numerical fits~\eqref{eq:fit-large-D}--\eqref{eq:fit-large-I} for incident proton momentum $50\text{ GeV}<P<450\text{ GeV}$.}
        \label{fig:rel-dev-large}
\end{figure}
\ref{fig:rel-dev-small},\,\ref{fig:rel-dev-large}.

\bibliographystyle{JHEP}
\bibliography{EM-FF-biblio.bib}

\providecommand{\href}[2]{#2}\begingroup\raggedright\begin{thebibliography}{10}

\bibitem{Gorbunov:2024}
D.~Gorbunov and E.~Kriukova, \emph{{Dark photon production via inelastic proton
  bremsstrahlung with Pauli form factor}}, {\emph{Phys.Part.Nucl.} {\bfseries
  56} (2025) } [\href{https://arxiv.org/abs/2409.11089}{{\ttfamily
  2409.11089}}].

\bibitem{Mosphys}
E.~Kriukova, ``{Dark photon production via inelastic proton bremsstrahlung}.''
  \url{https://mosphys.ru/indico/event/7/contributions/387/attachments/318/496/Mosphys.pdf},
  1 March 2024.

\bibitem{SessiyaOFNRAN}
E.~Kriukova, ``{Dark photon production via inelastic proton bremsstrahlung with
  Pauli form factor}.''
  \url{https://indico.jinr.ru/event/4174/contributions/25715/attachments/18792/32118/Kriukova-Sessiya-RAS.pdf},
  5 April 2024.

\bibitem{Agrawal:2021dbo}
P.~Agrawal et~al., \emph{{Feebly-interacting particles: FIPs 2020 workshop
  report}}, \href{https://doi.org/10.1140/epjc/s10052-021-09703-7}{\emph{Eur.
  Phys. J. C} {\bfseries 81} (2021) 1015}
  [\href{https://arxiv.org/abs/2102.12143}{{\ttfamily 2102.12143}}].

\bibitem{Beacham:2019nyx}
J.~Beacham et~al., \emph{{Physics Beyond Colliders at CERN: Beyond the Standard
  Model Working Group Report}},
  \href{https://doi.org/10.1088/1361-6471/ab4cd2}{\emph{J. Phys. G} {\bfseries
  47} (2020) 010501} [\href{https://arxiv.org/abs/1901.09966}{{\ttfamily
  1901.09966}}].

\bibitem{Okun:1982xi}
L.B.~Okun, \emph{{Limits of electrodynamics: paraphotons?}}, {\emph{Sov. Phys.
  JETP} {\bfseries 56} (1982) 502}.

\bibitem{Galison:1983pa}
P.~Galison and A.~Manohar, \emph{{TWO Z's OR NOT TWO Z's?}},
  \href{https://doi.org/10.1016/0370-2693(84)91161-4}{\emph{Phys. Lett. B}
  {\bfseries 136} (1984) 279}.

\bibitem{Holdom:1985ag}
B.~Holdom, \emph{{Two U(1)'s and Epsilon Charge Shifts}},
  \href{https://doi.org/10.1016/0370-2693(86)91377-8}{\emph{Phys. Lett. B}
  {\bfseries 166} (1986) 196}.

\bibitem{Ruegg:2003ps}
H.~Ruegg and M.~Ruiz-Altaba, \emph{{The Stueckelberg field}},
  \href{https://doi.org/10.1142/S0217751X04019755}{\emph{Int. J. Mod. Phys. A}
  {\bfseries 19} (2004) 3265}
  [\href{https://arxiv.org/abs/hep-th/0304245}{{\ttfamily hep-th/0304245}}].

\bibitem{Miller:2021ycl}
D.J.~Miller, \emph{{The physics of the dark photon: a primer}},
  \href{https://doi.org/10.1080/00107514.2021.1959647}{\emph{Contemp. Phys.}
  {\bfseries 62} (2021) 110}.

\bibitem{Pospelov:2007mp}
M.~Pospelov, A.~Ritz and M.B.~Voloshin, \emph{{Secluded WIMP Dark Matter}},
  \href{https://doi.org/10.1016/j.physletb.2008.02.052}{\emph{Phys. Lett. B}
  {\bfseries 662} (2008) 53} [\href{https://arxiv.org/abs/0711.4866}{{\ttfamily
  0711.4866}}].

\bibitem{Filippi:2020kii}
A.~Filippi and M.~De~Napoli, \emph{{Searching in the dark: the hunt for the
  dark photon}}, \href{https://doi.org/10.1016/j.revip.2020.100042}{\emph{Rev.
  Phys.} {\bfseries 5} (2020) 100042}
  [\href{https://arxiv.org/abs/2006.04640}{{\ttfamily 2006.04640}}].

\bibitem{deNiverville:2016rqh}
P.~deNiverville, C.-Y.~Chen, M.~Pospelov and A.~Ritz, \emph{{Light dark matter
  in neutrino beams: production modelling and scattering signatures at
  MiniBooNE, T2K and SHiP}},
  \href{https://doi.org/10.1103/PhysRevD.95.035006}{\emph{Phys. Rev. D}
  {\bfseries 95} (2017) 035006}
  [\href{https://arxiv.org/abs/1609.01770}{{\ttfamily 1609.01770}}].

\bibitem{Araki:2023xgb}
T.~Araki, K.~Asai, T.~Iizawa, H.~Otono, T.~Shimomura and Y.~Takubo, \emph{{New
  constraint on dark photon at T2K off-axis near detector}},
  \href{https://doi.org/10.1007/JHEP11(2023)056}{\emph{JHEP} {\bfseries 11}
  (2023) 056} [\href{https://arxiv.org/abs/2308.01565}{{\ttfamily
  2308.01565}}].

\bibitem{DUNE:2020fgq}
{\scshape DUNE} collaboration, \emph{{Prospects for beyond the Standard Model
  physics searches at the Deep Underground Neutrino Experiment}},
  \href{https://doi.org/10.1140/epjc/s10052-021-09007-w}{\emph{Eur. Phys. J. C}
  {\bfseries 81} (2021) 322}
  [\href{https://arxiv.org/abs/2008.12769}{{\ttfamily 2008.12769}}].

\bibitem{Breitbach:2021gvv}
M.~Breitbach, L.~Buonocore, C.~Frugiuele, J.~Kopp and L.~Mittnacht,
  \emph{{Searching for physics beyond the Standard Model in an off-axis DUNE
  near detector}}, \href{https://doi.org/10.1007/JHEP01(2022)048}{\emph{JHEP}
  {\bfseries 01} (2022) 048}
  [\href{https://arxiv.org/abs/2102.03383}{{\ttfamily 2102.03383}}].

\bibitem{Gorbunov:2014wqa}
D.~Gorbunov, A.~Makarov and I.~Timiryasov, \emph{{Decaying light particles in
  the SHiP experiment: Signal rate estimates for hidden photons}},
  \href{https://doi.org/10.1103/PhysRevD.91.035027}{\emph{Phys. Rev. D}
  {\bfseries 91} (2015) 035027}
  [\href{https://arxiv.org/abs/1411.4007}{{\ttfamily 1411.4007}}].

\bibitem{SHiP:2015vad}
{\scshape SHiP} collaboration, \emph{{A facility to Search for Hidden Particles
  (SHiP) at the CERN SPS}},  \href{https://arxiv.org/abs/1504.04956}{{\ttfamily
  1504.04956}}.

\bibitem{SHiP:2020vbd}
{\scshape SHiP} collaboration, \emph{{Sensitivity of the SHiP experiment to
  dark photons decaying to a pair of charged particles}},
  \href{https://doi.org/10.1140/epjc/s10052-021-09224-3}{\emph{Eur. Phys. J. C}
  {\bfseries 81} (2021) 451}
  [\href{https://arxiv.org/abs/2011.05115}{{\ttfamily 2011.05115}}].

\bibitem{Altarelli:1977zs}
G.~Altarelli and G.~Parisi, \emph{{Asymptotic Freedom in Parton Language}},
  \href{https://doi.org/10.1016/0550-3213(77)90384-4}{\emph{Nucl. Phys. B}
  {\bfseries 126} (1977) 298}.

\bibitem{Boiarska:2019jym}
I.~Boiarska, K.~Bondarenko, A.~Boyarsky, V.~Gorkavenko, M.~Ovchynnikov and
  A.~Sokolenko, \emph{{Phenomenology of GeV-scale scalar portal}},
  \href{https://doi.org/10.1007/JHEP11(2019)162}{\emph{JHEP} {\bfseries 11}
  (2019) 162} [\href{https://arxiv.org/abs/1904.10447}{{\ttfamily
  1904.10447}}].

\bibitem{Foroughi-Abari:2021zbm}
S.~Foroughi-Abari and A.~Ritz, \emph{{Dark sector production via proton
  bremsstrahlung}},
  \href{https://doi.org/10.1103/PhysRevD.105.095045}{\emph{Phys. Rev. D}
  {\bfseries 105} (2022) 095045}
  [\href{https://arxiv.org/abs/2108.05900}{{\ttfamily 2108.05900}}].

\bibitem{Likhoded:2010pc}
A.K.~Likhoded, A.V.~Luchinsky and A.A.~Novoselov, \emph{{Light hadron
  production in inclusive pp-scattering at LHC}},
  \href{https://doi.org/10.1103/PhysRevD.82.114006}{\emph{Phys. Rev. D}
  {\bfseries 82} (2010) 114006}
  [\href{https://arxiv.org/abs/1005.1827}{{\ttfamily 1005.1827}}].

\bibitem{Blumlein:2013cua}
J.~Bl\"umlein and J.~Brunner, \emph{{New Exclusion Limits on Dark Gauge Forces
  from Proton Bremsstrahlung in Beam-Dump Data}},
  \href{https://doi.org/10.1016/j.physletb.2014.02.029}{\emph{Phys. Lett. B}
  {\bfseries 731} (2014) 320}
  [\href{https://arxiv.org/abs/1311.3870}{{\ttfamily 1311.3870}}].

\bibitem{Kim:1973he}
K.J.~Kim and Y.-S.~Tsai, \emph{{Improved Weizsacker-Williams method and its
  application to lepton and W boson pair production}},
  \href{https://doi.org/10.1103/PhysRevD.8.3109}{\emph{Phys. Rev. D} {\bfseries
  8} (1973) 3109}.

\bibitem{Gorbunov:2023jnx}
D.~Gorbunov and E.~Kriukova, \emph{{Dark photon production via elastic proton
  bremsstrahlung with non-zero momentum transfer}},
  \href{https://doi.org/10.1007/JHEP01(2024)058}{\emph{JHEP} {\bfseries 01}
  (2024) 058} [\href{https://arxiv.org/abs/2306.15800}{{\ttfamily
  2306.15800}}].

\bibitem{Kriukova:2024wsi}
E.~Kriukova, \emph{{Dark photon emission in elastic proton bremsstrahlung}},
  \href{https://doi.org/10.22323/1.455.0011}{\emph{PoS} {\bfseries
  ICPPCRubakov2023} (2024) 011}
  [\href{https://arxiv.org/abs/2404.04704}{{\ttfamily 2404.04704}}].

\bibitem{JeffersonLabHallA:1999epl}
{\scshape Jefferson Lab Hall A} collaboration, \emph{{G(E(p)) / G(M(p)) ratio
  by polarization transfer in polarized e p ---\ensuremath{>} e polarized p}},
  \href{https://doi.org/10.1103/PhysRevLett.84.1398}{\emph{Phys. Rev. Lett.}
  {\bfseries 84} (2000) 1398}
  [\href{https://arxiv.org/abs/nucl-ex/9910005}{{\ttfamily nucl-ex/9910005}}].

\bibitem{JeffersonLabHallA:2001qqe}
{\scshape Jefferson Lab Hall A} collaboration, \emph{{Measurement of G(Ep) /
  G(Mp) in polarized-e p ---\ensuremath{>} e polarized-p to Q**2 =
  5.6-GeV**2}},
  \href{https://doi.org/10.1103/PhysRevLett.88.092301}{\emph{Phys. Rev. Lett.}
  {\bfseries 88} (2002) 092301}
  [\href{https://arxiv.org/abs/nucl-ex/0111010}{{\ttfamily nucl-ex/0111010}}].

\bibitem{Puckett:2011xg}
A.J.R.~Puckett et~al., \emph{{Final Analysis of Proton Form Factor Ratio Data
  at $\mathbf{Q^2 = 4.0}$, 4.8 and 5.6 GeV$\mathbf{^2}$}},
  \href{https://doi.org/10.1103/PhysRevC.85.045203}{\emph{Phys. Rev. C}
  {\bfseries 85} (2012) 045203}
  [\href{https://arxiv.org/abs/1102.5737}{{\ttfamily 1102.5737}}].

\bibitem{A1:2013fsc}
{\scshape A1} collaboration, \emph{{Electric and magnetic form factors of the
  proton}}, \href{https://doi.org/10.1103/PhysRevC.90.015206}{\emph{Phys. Rev.
  C} {\bfseries 90} (2014) 015206}
  [\href{https://arxiv.org/abs/1307.6227}{{\ttfamily 1307.6227}}].

\bibitem{BESIII:2021rqk}
{\scshape BESIII} collaboration, \emph{{Measurement of proton electromagnetic
  form factors in the time-like region using initial state radiation at
  BESIII}}, \href{https://doi.org/10.1016/j.physletb.2021.136328}{\emph{Phys.
  Lett. B} {\bfseries 817} (2021) 136328}
  [\href{https://arxiv.org/abs/2102.10337}{{\ttfamily 2102.10337}}].

\bibitem{BaBar:2013ves}
{\scshape BaBar} collaboration, \emph{{Study of $e^+e^- \to p \bar{p}$ via
  initial-state radiation at BABAR}},
  \href{https://doi.org/10.1103/PhysRevD.87.092005}{\emph{Phys. Rev. D}
  {\bfseries 87} (2013) 092005}
  [\href{https://arxiv.org/abs/1302.0055}{{\ttfamily 1302.0055}}].

\bibitem{CMD-3:2015fvi}
{\scshape CMD-3} collaboration, \emph{{Study of the process $e^+e^-\to
  p\bar{p}$ in the c.m. energy range from threshold to 2 GeV with the CMD-3
  detector}}, \href{https://doi.org/10.1016/j.physletb.2016.04.048}{\emph{Phys.
  Lett. B} {\bfseries 759} (2016) 634}
  [\href{https://arxiv.org/abs/1507.08013}{{\ttfamily 1507.08013}}].

\bibitem{CLEO:2005tiu}
{\scshape CLEO} collaboration, \emph{{Precision measurements of the timelike
  electromagnetic form-factors of pion, kaon, and proton}},
  \href{https://doi.org/10.1103/PhysRevLett.95.261803}{\emph{Phys. Rev. Lett.}
  {\bfseries 95} (2005) 261803}
  [\href{https://arxiv.org/abs/hep-ex/0510005}{{\ttfamily hep-ex/0510005}}].

\bibitem{Lin:2021xrc}
Y.-H.~Lin, H.-W.~Hammer and U.-G.~Mei\ss{}ner, \emph{{New Insights into the
  Nucleon\textquoteright{}s Electromagnetic Structure}},
  \href{https://doi.org/10.1103/PhysRevLett.128.052002}{\emph{Phys. Rev. Lett.}
  {\bfseries 128} (2022) 052002}
  [\href{https://arxiv.org/abs/2109.12961}{{\ttfamily 2109.12961}}].

\bibitem{Hoferichter:2016duk}
M.~Hoferichter, B.~Kubis, J.~Ruiz~de Elvira, H.W.~Hammer and U.G.~Mei\ss{}ner,
  \emph{{On the $\pi\pi$ continuum in the nucleon form factors and the proton
  radius puzzle}}, \href{https://doi.org/10.1140/epja/i2016-16331-7}{\emph{Eur.
  Phys. J. A} {\bfseries 52} (2016) 331}
  [\href{https://arxiv.org/abs/1609.06722}{{\ttfamily 1609.06722}}].

\bibitem{Lin:2021umz}
Y.-H.~Lin, H.-W.~Hammer and U.-G.~Mei\ss{}ner, \emph{{Dispersion-theoretical
  analysis of the electromagnetic form factors of the nucleon: Past, present
  and future}},
  \href{https://doi.org/10.1140/epja/s10050-021-00562-0}{\emph{Eur. Phys. J. A}
  {\bfseries 57} (2021) 255}
  [\href{https://arxiv.org/abs/2106.06357}{{\ttfamily 2106.06357}}].

\bibitem{Faessler:2009tn}
A.~Faessler, M.I.~Krivoruchenko and B.V.~Martemyanov, \emph{{Once more on
  electromagnetic form factors of nucleons in extended vector meson dominance
  model}}, \href{https://doi.org/10.1103/PhysRevC.82.038201}{\emph{Phys. Rev.
  C} {\bfseries 82} (2010) 038201}
  [\href{https://arxiv.org/abs/0910.5589}{{\ttfamily 0910.5589}}].

\bibitem{Aitchison:1972ay}
I.J.R.~Aitchison, \emph{{K-MATRIX FORMALISM FOR OVERLAPPING RESONANCES}},
  \href{https://doi.org/10.1016/0375-9474(72)90305-3}{\emph{Nucl. Phys. A}
  {\bfseries 189} (1972) 417}.

\bibitem{Dubnicka:2002yp}
S.~Dubnicka, A.-Z.~Dubnickova and P.~Weisenpacher, \emph{{Nucleon
  electromagnetic structure revisited}},
  \href{https://doi.org/10.1088/0954-3899/29/2/316}{\emph{J. Phys. G}
  {\bfseries 29} (2003) 405}
  [\href{https://arxiv.org/abs/hep-ph/0208051}{{\ttfamily hep-ph/0208051}}].

\bibitem{Adamuscin:2016rer}
C.~Adamu\v{s}\v{c}in, E.~Barto\v{s}, S.~Dubni\v{c}ka and A.Z.~Dubni\v{c}kov\'a,
  \emph{{Numerical values of $f^F$, $f^D$, $f^S$ coupling constants in $SU(3)$
  invariant interaction Lagrangian of vector-meson nonet with $1/2^+$ octet
  baryons}}, \href{https://doi.org/10.1103/PhysRevC.93.055208}{\emph{Phys. Rev.
  C} {\bfseries 93} (2016) 055208}
  [\href{https://arxiv.org/abs/1601.06190}{{\ttfamily 1601.06190}}].

\bibitem{Dubnickova:2020heq}
A.Z.~Dubnickova and S.~Dubnicka, \emph{{Proton em form factors data are in
  disagreement with new $\sigma_{tot}(e^+e^- \to p\bar p)$ measurements}},
  \href{https://arxiv.org/abs/2010.15872}{{\ttfamily 2010.15872}}.

\bibitem{Haberzettl:2011zr}
H.~Haberzettl, F.~Huang and K.~Nakayama, \emph{{Dressing the electromagnetic
  nucleon current}},
  \href{https://doi.org/10.1103/PhysRevC.83.065502}{\emph{Phys. Rev. C}
  {\bfseries 83} (2011) 065502}
  [\href{https://arxiv.org/abs/1103.2065}{{\ttfamily 1103.2065}}].

\bibitem{Choi:2019nvk}
H.-M.~Choi, T.~Frederico, C.-R.~Ji and J.P.B.C.~de~Melo, \emph{{Pion off-shell
  electromagnetic form factors: data extraction and model analysis}},
  \href{https://doi.org/10.1103/PhysRevD.100.116020}{\emph{Phys. Rev. D}
  {\bfseries 100} (2019) 116020}
  [\href{https://arxiv.org/abs/1908.01185}{{\ttfamily 1908.01185}}].

\bibitem{Leao:2024agy}
J.~Le\~ao, J.P.B.C.~de~Melo, T.~Frederico, H.-M.~Choi and C.-R.~Ji,
  \emph{{Off-shell pion properties: electromagnetic form factors and
  light-front wave functions}},
  \href{https://arxiv.org/abs/2406.07743}{{\ttfamily 2406.07743}}.

\bibitem{Feuster:1998cj}
T.~Feuster and U.~Mosel, \emph{{Photon and meson induced reactions on the
  nucleon}}, \href{https://doi.org/10.1103/PhysRevC.59.460}{\emph{Phys. Rev. C}
  {\bfseries 59} (1999) 460}
  [\href{https://arxiv.org/abs/nucl-th/9803057}{{\ttfamily nucl-th/9803057}}].

\bibitem{Alterelli:1964bu}
G.~Altarelli and F.~Buccella, \emph{{Single photon emission in high-energy
  e+-e- collisions}}, \href{https://doi.org/10.1007/BF02748859}{\emph{Nuovo
  Cim} {\bfseries 34} (1964) 1337}.

\bibitem{Baier:1966jf}
V.N.~Baier, V.S.~Fadin and V.A.~Khoze, \emph{{Photon bremsstrahlung in
  collisions of high-energy electrons}}, {\emph{Sov. Phys. JETP} {\bfseries 24}
  (1966) 760}.

\bibitem{BERESTETSKII1982354}
V.~Berestetskii, E.~Lifshitz and L.~Pitaevskii, \emph{Interaction of electrons
  with photons},  in \emph{Quantum Electrodynamics (2nd edition)}, (Oxford),
  pp.~354--455, Butterworth-Heinemann (1982),
  \href{https://doi.org/10.1016/B978-0-08-050346-2.50016-7}{DOI}.

\bibitem{Berryman:2019dme}
J.M.~Berryman, A.~de~Gouvea, P.J.~Fox, B.J.~Kayser, K.J.~Kelly and J.L.~Raaf,
  \emph{{Searches for Decays of New Particles in the DUNE Multi-Purpose Near
  Detector}}, \href{https://doi.org/10.1007/JHEP02(2020)174}{\emph{JHEP}
  {\bfseries 02} (2020) 174}
  [\href{https://arxiv.org/abs/1912.07622}{{\ttfamily 1912.07622}}].

\bibitem{Miano:2020gjq}
A.~Miano, A.~Fiorillo, A.~Salzano, A.~Prota and R.~Jacobsson, \emph{{The
  structural design of the decay volume for the Search for Hidden Particles
  (SHIP) project}},
  \href{https://doi.org/10.1007/s43452-020-00152-9}{\emph{Archives of Civil and
  Mechanical Engineering} {\bfseries 21} (2020) 3}.

\bibitem{ParticleDataGroup:2022pth}
{\scshape Particle Data Group} collaboration, \emph{{Review of Particle
  Physics}}, \href{https://doi.org/10.1093/ptep/ptac097}{\emph{PTEP} {\bfseries
  2022} (2022) 083C01}.

\bibitem{PDG:Materials}
D.~Groom, ``{Atomic and Nuclear Properties of Materials}.''
  \url{https://pdg.lbl.gov/2023/AtomicNuclearProperties/}, 2023.

\bibitem{Denisov:1971jb}
S.P.~Denisov, S.V.~Donskov, Y.P.~Gorin, A.I.~Petrukhin, Y.D.~Prokoshkin,
  D.A.~Soyanova et~al., \emph{{Total cross-sections of pi+, K+ and p on protons
  and deuterons in the momentum range 15-GeV/c to 60-GeV/c}},
  \href{https://doi.org/10.1016/0370-2693(71)90739-8}{\emph{Phys. Lett. B}
  {\bfseries 36} (1971) 415}.

\bibitem{Carroll:1975xf}
A.S.~Carroll et~al., \emph{{Total Cross-Sections of $\pi^\pm$, $K^\pm$, $p$,
  and $\bar{p}$ on Protons and Deuterons Between 23-GeV/c and 280-GeV/c}},
  \href{https://doi.org/10.1016/0370-2693(76)90155-6}{\emph{Phys. Lett. B}
  {\bfseries 61} (1976) 303}.

\bibitem{Denisov:1973zv}
S.P.~Denisov, S.V.~Donskov, Y.P.~Gorin, R.N.~Krasnokutsky, A.I.~Petrukhin,
  Y.D.~Prokoshkin et~al., \emph{{Absorption cross-sections for pions, kaons,
  protons and anti-protons on complex nuclei in the 6-GeV/c to 60-GeV/c
  momentum range}},
  \href{https://doi.org/10.1016/0550-3213(73)90351-9}{\emph{Nucl. Phys. B}
  {\bfseries 61} (1973) 62}.

\bibitem{Carroll:1978hc}
A.S.~Carroll et~al., \emph{{Absorption Cross-Sections of $\pi^{\pm}$,
  $K^{\pm}$, p and $\bar{p}$ on Nuclei Between 60 GeV/c and 280 GeV/c}},
  \href{https://doi.org/10.1016/0370-2693(79)90226-0}{\emph{Phys. Lett. B}
  {\bfseries 80} (1979) 319}.

\bibitem{PANDA:2021ozp}
{\scshape PANDA} collaboration, \emph{{PANDA Phase One}},
  \href{https://doi.org/10.1140/epja/s10050-021-00475-y}{\emph{Eur. Phys. J. A}
  {\bfseries 57} (2021) 184}
  [\href{https://arxiv.org/abs/2101.11877}{{\ttfamily 2101.11877}}].

\bibitem{Kuraev:2012ca}
E.A.~Kuraev, Y.M.~Bystritskiy, V.V.~Bytev, E.~Tomasi-Gustafsson, A.~Dbeyssi and
  E.~Tomasi-Gustafsson, \emph{{Annihilation of $\bar p+p\to e^++e^-+ \pi^0$ and
  $\bar p+p\to \gamma + \pi^0$ through $\omega$-meson intermediate state}},
  \href{https://doi.org/10.1134/S1063776112060064}{\emph{J. Exp. Theor. Phys.}
  {\bfseries 115} (2012) 93} [\href{https://arxiv.org/abs/1012.5720}{{\ttfamily
  1012.5720}}].

\bibitem{Adamuscin:2006dfq}
C.~Adamuscin, E.A.~Kuraev, E.~Tomasi-Gustafsson and F.E.~Maas, \emph{{Testing
  axial and electromagnetic nucleon form factors in time-like regions in the
  processes anti-p + n ---\ensuremath{>} pi- + l- + l+ and anti-p + p
  ---\ensuremath{>} pi0 + l- + l+, l=e, mu}},
  \href{https://doi.org/10.1103/PhysRevC.75.045205}{\emph{Phys. Rev. C}
  {\bfseries 75} (2007) 045205}
  [\href{https://arxiv.org/abs/hep-ph/0610429}{{\ttfamily hep-ph/0610429}}].

\bibitem{Ovchynnikov:2023cry}
M.~Ovchynnikov, J.-L.~Tastet, O.~Mikulenko and K.~Bondarenko,
  \emph{{Sensitivities to feebly interacting particles: Public and unified
  calculations}},
  \href{https://doi.org/10.1103/PhysRevD.108.075028}{\emph{Phys. Rev. D}
  {\bfseries 108} (2023) 075028}
  [\href{https://arxiv.org/abs/2305.13383}{{\ttfamily 2305.13383}}].

\bibitem{Du:2022hms}
M.~Du, R.~Fang and Z.~Liu, \emph{{Millicharged particles from proton
  bremsstrahlung in the atmosphere}},
  \href{https://arxiv.org/abs/2211.11469}{{\ttfamily 2211.11469}}.

\bibitem{Wu:2024iqm}
H.~Wu, E.~Hardy and N.~Song, \emph{{Searching for heavy millicharged particles
  from the atmosphere}},  \href{https://arxiv.org/abs/2406.01668}{{\ttfamily
  2406.01668}}.

\bibitem{Kondratyuk:2005kk}
S.~Kondratyuk, P.G.~Blunden, W.~Melnitchouk and J.A.~Tjon, \emph{{Delta
  resonance contribution to two-photon exchange in electron-proton
  scattering}},
  \href{https://doi.org/10.1103/PhysRevLett.95.172503}{\emph{Phys. Rev. Lett.}
  {\bfseries 95} (2005) 172503}
  [\href{https://arxiv.org/abs/nucl-th/0506026}{{\ttfamily nucl-th/0506026}}].

\bibitem{Foroughi-Abari:2024xlj}
S.~Foroughi-Abari, P.~Reimitz and A.~Ritz, \emph{{A Closer Look at Dark Vector
  Splitting Functions in Proton Bremsstrahlung}},
  \href{https://arxiv.org/abs/2409.09123}{{\ttfamily 2409.09123}}.

\end{thebibliography}\endgroup

\end{document}